\documentclass[11pt]{article}
\usepackage[utf8]{inputenc}
\usepackage{graphicx}
\usepackage{subcaption}
\usepackage{amsthm}
\usepackage{mathrsfs}
\usepackage{slashed}
\usepackage{float}

\begin{document}
\title{\bf Role of dynamical early dark energy in Hubble tension through warm inflation}
\vspace{-5mm}
\date{}
\maketitle
\begin{center}
\large{Anupama B\footnote[1]{21phph19@uohyd.ac.in (corresponding author)}}\\
\small{School of Physics, University of Hyderabad, Hyderabad, 500046, India.} 
\end{center}
\begin{abstract}
The role of dissipation from the thermal bath of minimal warm inflation (MWI)is examined with the Cosmic Microwave Background (CMB) to assess its ability to mimic a dynamical dark energy component. An increase in the dissipation strength modifies temperature anisotropies and shifts the phase and peak structure of the CMB TT angular power spectrum in a manner analogous to evolving dark energy. Bayesian parameter estimation for cosmological parameters of MWI using Markov Chain Monte Carlo analysis with \texttt{Planck18 +  BICEP/Keck 2018 + sn.pantheon + SH0ES}  likelihoods reveals deviations from standard $\Lambda$CDM constraints, accompanied by an enhancement in the matter power spectrum that signals possible physics beyond the conventional model. Notably, the present day Hubble parameter increases with dissipation strength, indicating that minimal warm inflation can naturally generate a dynamical early dark energy like behavior capable of alleviating the Hubble tension.
Within this unified framework, early and late time estimates of $H_0$ can be reconciled. Further, the time varying dark energy equation of state consistent with recent DESI DR2 indication favours a quantum gravity phase preceding the inflationary epoch. \\
 \vspace{1pc}
\end{abstract}
\vspace{1pc}
\section{Introduction}
Resolving the major issues in standard cosmology requires a natural and suitable inflationary model that is consistent with the ongoing observations.  Warm inflation (WI) assumes a non negligible radiation during the beginning of inflation and the subsequent interaction of the inflaton with other degrees of freedom, releasing the decay products into a persistent thermal bath \cite{AB,ab2,ab3}. WI has the built-in ability to describe the dynamics of inflation and the following stages by accommodating standard model particle physics interactions \cite{AB,RD,story,Adev,cqg2,bridge,rudnei}. The dissipative dynamics in WI determine the nature of the interaction between different fields. Conventional inflationary models, such as chaotic inflation and attractor models, are found to outperform the cold inflationary (CI) scenario in contrast with the multi-field warm inflationary setup \cite{cqg2, chaotic, Suratna,out1,out2,out3,out4}.  Keeping this in view, efforts have been made in warm inflationary model building to incorporate elements like supersymmetry and axionic interactions that aid in minimising thermal mass corrections. The Minimal Warm Inflation (MWI) \cite{kim,sph,chetan} is considered a state-of-the-art model \cite{rudnei, soam} in the strong dissipation regime of warm inflationary scenario, where the inflaton assumes axionic coupling to non-Abelian gauge fields. Shift symmetry of the axion can effortlessly prevent any backreaction and the related huge thermal corrections to the inflaton mass.\\
\\
The results from Dark Energy Spectroscopic Instrument (DESI) DR2, suggesting a varying equation of state for the dark energy, have received great attention in recent times \cite{desi}. In view of this, a finding on alleviating the Hubble tension \cite{cqg2} with dynamical dark energy like behaviour from WI in the early universe highlights the potential of interactions and dissipation during WI to increase the present value of the Hubble parameter ($H_0$).  If this is true, then the dynamical dark energy from WI must interfere with the leading interactions of the photon-baryon acoustic oscillations before the recombination, which may lead to a variation in the phase or amplitude of the angular power spectrum of CMB \cite{waynehu, wh}. Any variation in the spatial temperature due to the dissipation in the thermal bath of WI may contribute to the CMB temperature anisotropies. The aforementioned reasons compel us to explore WI and the dynamical nature of dark energy in the early universe through CMB. This paper unveils new theoretical evidence that validates the dynamical dark energy arising naturally from the warm inflationary dynamics through its signatures on the TT mode angular power spectrum of CMB and the matter power spectrum. WI may have a significant edge over the CI by accommodating dynamical dark energy that may reduce the Hubble tension, thus indicating the necessity to revamp the cosmological parameters in the $\Lambda$CDM concordance model, which is currently under intense scrutiny. MCMC method is used to analyse the performance of MWI model in the context of dynamical dark energy and Hubble tension.\\
\\The role of Early Dark Energy (EDE) from axionic fields in resolving the Hubble tension is a widely discussed topic in cosmology \cite{kami, kami1, kami2, ede1, ede2, ede3, ede4, ede5}. 
The core idea of EDE is to introduce a transient exotic component that contributes to the total energy density shortly before recombination and then dilutes away \cite{m1, m2, m3, m4, m5, m6, m7}. This auxiliary energy may boost the inferred value of $H_0$. In slow-roll realisations, the required friction can be generated by coupling an axion-like field to a thermal bath in non-Abelian sectors \cite{wi1, wi2, wi3, wi4, wi5}. Relatedly, scenarios with interacting thermal baths can employ a mediator particle that can transfer its entropy into lighter species, raising the effective number of relativistic degrees of freedom ($g_*$). This mechanism raises the pre-recombination energy density and avoids the problems that a simple increase in $g_*$ would introduce in CMB. The present work is motivated by the choice of MWI model, which focuses on the interaction between the axion ($\Phi$) and an arbitrary Yang-Mills group ($\mathcal{G}^a_{\mu\nu}$) given by the Lagrangian  \cite{kim,sph,chetan}
\begin{equation}
\mathcal{L}_{int} = \frac{\alpha}{16 \pi} \frac{\Phi}{f} \tilde \mathcal{G}^{\mu \nu}_a \mathcal{G}^a_{\mu \nu}, \label{lgm}
\end{equation}
with the dissipation coefficient ($\gamma$) having a cubic relation with the temperature ($T$) of the thermal bath as,
\begin{eqnarray}
\gamma(T)=\kappa(\alpha,N_c,N_f)\frac{\alpha^5 T^3}{f^2}.
\end{eqnarray}
Here, $f$ is the axion decay constant,  $\alpha = \frac{g^2}{4\pi}$ represents the Yang-Mills gauge coupling, and $\kappa$ is a dimensionless quantity that depends on the number of colours ($N_c$) and flavours ($N_f$) of the gauge group. In the current work, a quadratic inflaton potential $V(\Phi) = \frac{1}{2} m^2\Phi^2$, is considered in a slow-roll setting. Such a simple potential with a bare mass ($m$) term does not alter the background interactions and therefore is crucial in preserving the underlying physics. Despite its conceptual simplicity, EDE acts like a time dependent fluid in a relativistic theory, which supports spatial perturbations. Therefore, EDE models must also account for possible density fluctuations in the EDE component.
In addition to the homogeneous background component $\Phi(t)$ called inflaton, the scalar field ($\Phi(x,t)$) also exhibits small perturbations ($\delta \Phi(x,t)$), which can be expressed as
\begin{equation}
\Phi(x,t) = \Phi(t) + \delta \Phi(x,t). 
\end{equation}
In the cold inflationary scenario, the perturbations of an arbitrary massive scalar field arise from quantum fluctuations, and their amplitude and variance can be derived from\begin{equation}
\ddot \delta  \Phi(x,t)_{CI} + [k^2 + m^2]  \delta  \Phi(x,t)_{CI} = 0 
\end{equation}
as
\begin{equation}\label{bj90}
<|\delta \Phi|^2_{CI} > \  \ \simeq \  \frac{3H^4}{8\pi^2m^2} \bigg[ 1 - e^{- \frac{2m^2N}{3H^2}} \bigg].
\end{equation}
Here, $H$ denotes the Hubble parameter during inflation,  $k$ represents the comoving wavenumber of the mode crossing the horizon, $m$ is the mass of the scalar field $\Phi(t)$, and $N$ corresponds to the number of e-folds, which quantifies the duration of the inflationary phase.
In warm inflation, the equation of motion of inflaton and hence the fluctuations have a thermal contribution, which takes the form, 
\begin{eqnarray}
& \ddot \Phi + 3H (1+Q) \dot \Phi + V'(\Phi) = 0, \label{motion}\\
& \delta \ddot  \Phi(k,t)_{WI} + (3H+ \gamma) \delta \dot \Phi(k,t)_{WI} +  (\frac{k}{a^2}+m^2)\delta\Phi(k,t)_{WI} = \xi(k,t),\\
& <\delta \Phi^2_{WI}> \  \sim
\cases{
 \frac{3HT}{4\pi} & for WI with $Q<<1$ \cr 
\frac{\sqrt{\gamma H} T}{2\pi^2} &  for WI with $Q>>1$, \cr } \label{fluct}
\end{eqnarray} 
where, $Q=\frac{\gamma}{3H}$represents the ratio of dissipation to expansion and $\xi$ denotes the thermal noise fluctuation, whose ensemble average is given in terms of scale factor ($a$) by
\begin{eqnarray}
<\xi(x,t)\xi(x',t)> \ = 2\gamma T a^{-3} \delta(x-x')\delta(t-t').
\end{eqnarray}
In the early universe, quantum fluctuations can dominate over the homogeneous background component of the field. Therefore, the inflationary potential in MWI with $Q>>1$ can be rewritten  using Eq. (\ref{fluct}),
\begin{equation}\label{V}
V(\Phi) \  \simeq \ \frac{1}{2}m^2  <|\delta \Phi|^2_{WI} > \ \simeq \ \frac{m^2 \sqrt{\gamma H}T}{4 \pi^2} .
\end{equation}
The pressure and energy density for inflaton in a flat Friedmann-Lemaître-Robertson-Walker (FLRW) metric,  $ds^2 =- dt^2 + a(t)^2(dr^2 + r^2 (d\theta^2+\sin^2\theta d\varphi^2))$ is,
\begin{eqnarray}
P = \frac{\dot \Phi^2}{2} - V(\Phi), \label{pressure} \\
\rho=\frac{\dot \Phi^2}{2} +V(\Phi) .
\end{eqnarray}
The barotropic fluid equation relates the pressure ($P$) and energy density ($\rho$) of the early universe through the equation of state ($\omega$),
\begin{equation}
P =\omega \rho.
\end{equation}
Substituting for $\rho$, the Friedmann equation can be rewritten in terms of $P$ and $\omega$, 
\begin{eqnarray} \label{fried}
H^2 \  = \frac{\rho}{3m_{pl}^2} = \frac{P}{3 \omega m_{pl}^2}.
\end{eqnarray}
The Hubble parameter can be expressed interms of the energy densities of various components in the universe as,
\begin{equation}
H^2 = (\Omega_m+\Omega_r+\Omega_\Lambda )H_0^2.
\end{equation}
During the onset of inflation $\Omega_m+\Omega_r=0$.
Therefore, \begin{equation}\label{sc}
H^2 = \Omega_\Lambda  H_0^2.
\end{equation}
Here, it is assumed that the early dark energy component gradually evolves into the present day cosmological constant ($\Lambda$), contributing a fractional energy density $\Omega_\Lambda$.
The pressure can be rewritten in terms of $H_0$ from the Friedmann equation as
\begin{equation}\label{newp}
P = 3 \omega m_{pl}^2H^2=3 \omega m_{pl}^2\Omega_\Lambda H_0^2.
\end{equation}
The energy density of the inflaton in WI, 
\begin{eqnarray}\label{rhotemp}
\rho = \frac{\pi^2}{30}g_*T^4,
\end{eqnarray}
is related to the temperature of the thermal bath ($T$) and the number of relativistic degrees of freedom ($g_*$).
An accelerated expansion during inflation requires the slow roll condition ($\frac{\dot\Phi^2}{2}<<V$). Therefore, in a slow roll regime, Eq. (\ref{motion}) reduces to
\begin{eqnarray}\label{eqwarm}
V' = -3H(1+Q)\dot\Phi .
\end{eqnarray}
Using Eqs. (\ref{fried}), (\ref{rhotemp} and (\ref{eqwarm}), the thermal contribution from warm inflation can be computed,
\begin{eqnarray}\label{TH}
\frac{T}{H} = c \bigg[\frac{Q}{(1+Q)^2}\bigg]^{\frac{1}{4}},
\end{eqnarray} where $c = \bigg( \frac{135}{64} \frac{\epsilon^{CI}}{\pi^4 V g_*} \bigg)^{\frac{1}{4}}$.
The slow roll parameters in CI are defined as,
\begin{eqnarray}
\epsilon^{CI}=\epsilon_H&=& \frac{-\dot H}{H^2} \simeq \frac{m_{pl}^2}{2}\bigg(\frac{V'(\Phi)}{V(\Phi)}\bigg)^2 \simeq \epsilon_v, \\ 
\eta^{CI}=\eta_H &=& \frac{- \ddot \Phi}{H \dot \Phi} \simeq m_{pl}^2 \frac{V''(\Phi)}{V(\Phi)} \simeq \eta_v.
\end{eqnarray} 
Whereas, in WI, 
\begin{eqnarray}
& \epsilon^{WI}= \epsilon_H =  \frac{\epsilon_v}{1+Q}, \label{epwarm} \\ 
& \eta^{WI}= \eta_H =  \frac{1}{1+Q}(\eta_v - \beta + \frac{\beta -\eta_v}{1+Q}).
\end{eqnarray}
The corresponding first ($\epsilon^{WI}$) and second ($\eta^{WI}$) slow roll parameters along with the scalar spectral index ($n_s^{WI}$) and the tensor to scalar ratio ($r^{WI}$)  for axionic MWI with $Q>>1$ can be expressed in terms of $Q$ and the e-folding number ($N$),
\begin{eqnarray}
\epsilon^{WI}  &=& \frac{1}{2NQ},\label{cif1}\\
\eta^{WI} &=& \frac{1}{2NQ}+3,\\
n_s^{WI} &=& 1- \frac{3}{8QN} + \frac{2QG'_{l=3}(Q)}{7NG_{l=3}(Q)},\\
r^{WI} &=& \frac{8}{ Nc \sqrt{3\pi}Q^\frac{1}{4}G_{l=3}(Q)} \label{cif4},
\end{eqnarray}
where, $G_{l=3}(Q) =  1+4.981Q^{1.946}+0.127Q^{4.336}$ \cite{RD}.
Comparing Eqs. (\ref{cif1}-\ref{cif4}) with recent CMB data suggests a valid range $1<Q<8$ for MWI with quadratic inflaton potential \cite{cqg2}. $Q>1$ indicates that the underlying model lies in the strong dissipation regime. The focal point of this paper is the viable and self consistent explanation for dynamical early dark energy from the multifield MWI that can simultaneously address the Hubble tension. The effect of strong dissipation from MWI in the early universe on CMB is studied for dynamical dark energy like behaviour in the next section and its subsections. Its implications on the $\Lambda$CDM model parameters are investigated. If WI naturally encompasses dynamical dark energy, then it may open a new direction for explaining the increased expansion rate in the early universe when compared with $Planck$ 2018 data. Thus, the outcome of the study may play a role in alleviating some of the persistent tensions in cosmology, especially the Hubble tension. The conclusions are presented in section (\ref{conc}).
\section{ Evidence for dynamical dark energy from WI}
Following the remarkable discovery that the universe is currently experiencing a second phase of accelerated expansion attributed to an unknown component referred to as dark energy, numerous efforts have been made to construct a theoretical framework that accounts for both the early time inflationary epoch and the present accelerated expansion.
In the current work, a warm inflationary scenario is considered in which an axionic field acts as the inflaton. In such a setting, a pre-existing radiation bath with a finite energy density is sustained throughout inflation, as it continuously receives energy from the inflaton field. The transfer of energy between the inflaton and radiation is governed by dissipation coefficients, which are assumed to be temperature dependent. At some stage, the radiation energy density surpasses that of the inflaton field, signalling the termination of the inflationary phase. As the universe undergoes a smooth transition to radiation domination, the dissipation coefficients decrease and rapidly become insignificant.
After the radiation dominated epoch,  the primordial nucleosynthesis, cold dark matter, together with baryonic matter governs the dynamics of the universe. This transition ensures a prolonged matter dominated era, providing sufficient time for the formation of the large scale structures observed today. Eventually, the scalar field 
identified with dark energy becomes the dominant energy component, driving the present phase of accelerated expansion. Generally, a strong dissipation regime is adopted for WI to improve observational viability, yielding a smaller tensor-to-scalar ratio and a spectral blue tilt, when compared to conventional cold inflation \cite{cqg2}. Deriving robust predictions for dynamical EDE from MWI and assessing its consistency with observational effects is not straight forward. A coordinated suit of existing approaches must be implemented systematically to analyse the appropriate dynamical dark energy effects from MWI.\\
\\
The standard model of cosmology consists of cold dark matter, cosmological constant, and standard model interactions governing baryons and photons.  Dynamical dark energy points toward the need for new physics beyond the standard cosmological model. In this work, an EDE scenario is implemented using a single scalar field (axion) that undergoes a transition from a slow-roll phase (the EDE phase) to an oscillatory phase (the decay phase). The energy density of EDE behaves like a dynamical dark energy until the end of inflation, after which it dilutes rapidly. Section (\ref{TTsection}) studies the fractional energy contribution from axionic EDE to the temperature anisotropies of CMB manifested in the TT mode angular power spectrum of CMB. In section (\ref{likeli}), the effect of axionic dynamical EDE from the MWI on the cosmological parameters is explored using Bayesian analysis using the MCMC method. The matter power spectrum is analysed in section (\ref{mps}) to find supporting evidence for the extra fractional energy contribution from EDE. Finally, in section (\ref{ddeht}), the dynamical EDE is proposed as an early universe workable solution to the Hubble tension.
\subsection{TT mode angular power spectrum of CMB}\label{TTsection} 
\begin{figure}[h]
\begin{center}
\includegraphics[width=\textwidth]{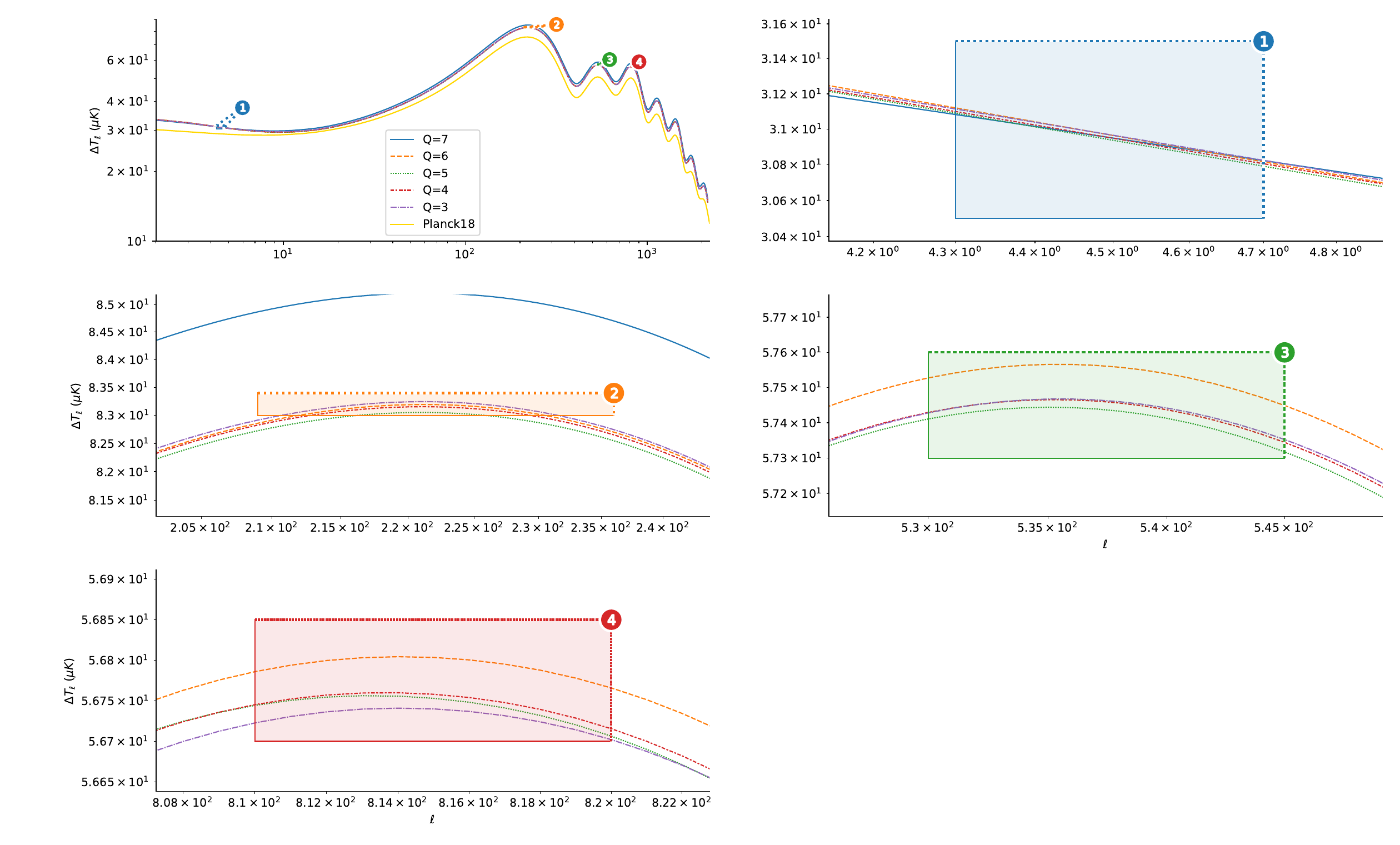}
\caption{TT mode angular power spectrum of CMB with various allowed dissipation strength ($Q$) for MWI (top left). The remaining figures represent the different zoomed regions of the angular power spectrum.
Here,  ($\Delta T_l )^2 = \frac{l(l+1)C_l }{2\pi}$ where $C_l$ the power spectrum of the multipole moments ($l$). }
\label{default}
\end{center}
\end{figure}

The core idea of a warm inflationary setting is to ensure a smooth transition into the radiation dominated era soon after inflation without the requirement of a separate reheating stage. But the continuous dissipation into the thermal bath during inflation may also contribute to the CMB temperature anisotropy, or in other words, any modification in the cosmological inflationary observables like $n_s^{WI}$ and $r^{WI}$ from the thermal bath of WI must reflect on the TT mode angular power spectrum of CMB. Using Eqs.(\ref{cif1}-\ref{cif4}) in the Code for Anisotropies in the Microwave Background (\texttt{CAMB})  \cite{CAMB} with $N=60$, the theoretical TT mode angular power spectrum of CMB is obtained and studied for MWI in the allowed range of $Q$. The COBE normalised results are presented in Fig. \ref{default}.
At a glance, the various angular power spectrum for MWI corresponding to different strengths of dissipation ($Q$) seems to overlap each other completely. But, a close scrutiny of the regions 1 to 4 in Fig. \ref{default} reveals the effect of MWI on the temperature anisotropies. 
In region 1, the power spectrum corresponding to $Q=7$ initially starts with a lower amplitude when compared with the power spectrum of other $Q$ values, gradually gains amplitude and rises above the rest of the $Q$. Similarly, a careful analysis shows that the TT mode angular power spectrum of MWI for $Q=6$ begins with a higher amplitude and eventually goes below that of  $Q=7$ and $Q=3$. Regions 2, 3 and 4 represent the first, second and third acoustic peaks of the CMB for MWI, respectively. The temperature difference ($\Delta T_l$) in CMB for $Q=3$ dominates over that of $Q=6$ in the first peak (see region 2), which gets reversed in the second peak (see region 3), i.e., the odd peak is higher than the even peak. Moreover, the amplitudes corresponding to $Q=3$, $Q=4$ and $Q=5$ in region 3 tend to overlap at lower multipoles when compared with those at higher multipoles, suggesting a phase shift. A phase shift is also observed in region 4, where $Q=4$ and $Q=5$ overlap at lower multipoles and the phase shift becomes higher at higher multipoles. Similarly, $Q=3$ and $Q=5$ diverge at $l=808$ and converge at $l=822$.  This shift in the peak and phase of the angular power spectrum of CMB for various allowed Q values of MWI resembles the varying dark energy equation of state ($\omega$) as pointed out by \cite{wh}. Although it was expected that the amplitude of TT mode angular power spectrum of CMB for $Q=3$ would dominate over $Q=6$ in region 4, this was not the case. This may be due to the existence of many free model parameters. Also, the amplitude of the angular power spectrum for all $Q$ in MWI has a higher amplitude when compared with the $Planck$ 2018 data, suggesting a modification to the energy budget of the universe. This is the first result of this paper that hints at dynamical dark energy behaviour in MWI, and it further motivates us to investigate the effect of dissipation and hence dynamical dark energy from WI on the $\Lambda$CDM cosmological model parameters.\\
\subsection{Bayesian parameter estimation using MCMC} \label{likeli}
\begin{figure}[h]
\begin{center}
\includegraphics[width=0.6\linewidth]{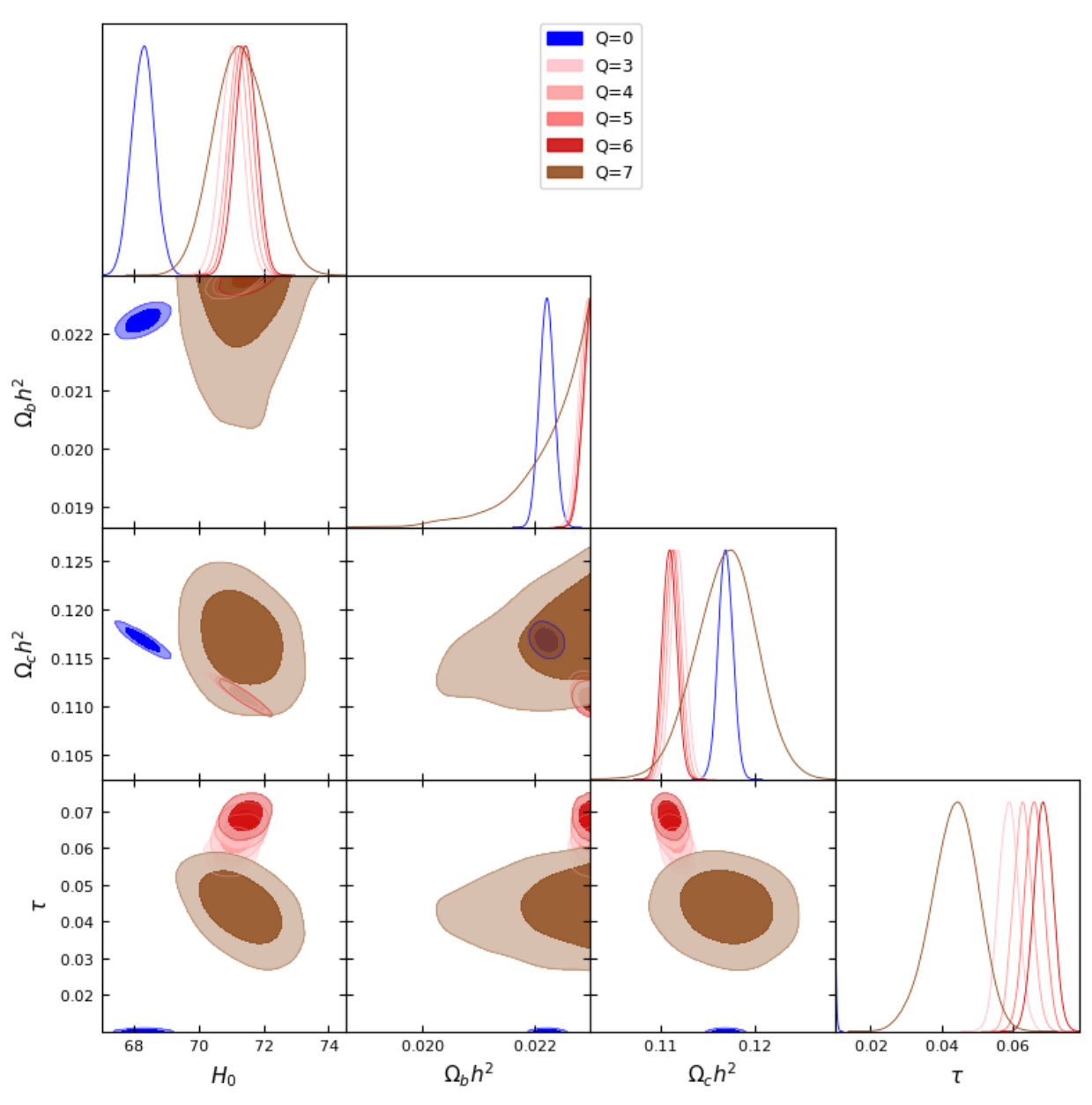}
\caption{Comparison between marginalised $68\%$ and $95\%$ confidence contours of different pairs of cosmological parameters in CI and MWI for various dissipation strengths obtained with baseline likelihoods from \texttt{Planck18 + lowTT + lowEE + highl CamSpec TTTEEE  +  BICEP/Keck 2018 + sn.pantheon + SH0ES} datasets.}
\label{mcmc}
\end{center}
\end{figure}

The well known $\Lambda$CDM standard cosmological model assumes a spatially flat FLRW universe with the cosmic energy budget dominated by dark energy and cold dark matter. Remarkably, despite involving only six parameters, $\Lambda$CDM successfully accounts for a wide range of cosmological observations. These parameters include the baryon ($\Omega_b h^2$) and cold dark matter ($\Omega_c h^2$) energy densities, which determine the matter content of the universe; the angular scale of the sound horizon at last scattering ($\Theta_{MC}$); the optical depth to reionization ($\tau$) which encodes information about the reionization epoch; and the amplitude ($A_s$) and spectral tilt ($n_s$) of the primordial scalar power spectrum. If dynamical dark energy is present in MWI, then it may modify some of the cosmological parameters. Therefore, it is necessary to verify the parameter space of the standard model by carrying out likelihood analysis in the framework of WI. Further, statistical analysis may strengthen the significance of the current results, thus recommending revisions to the $\Lambda$CDM model.
For a given scenario, one can predict observable quantities by collecting the corresponding observational data and comparing them to infer the values of the model parameters. These parameters cannot be determined exactly owing to unavoidable experimental uncertainties. Bayesian inference offers a systematic approach to parameter estimation by propagating measurement errors into probabilistic confidence levels. It is expressed through Bayes’ theorem as,
\begin{equation}
P(\theta | D)=\frac{P(D|\theta) P(\theta) }{P(D)}
\end{equation}
where, $P(\theta$) represents prior knowledge of the parameters $\theta$, $P(D|\theta)$ is the likelihood of obtaining the data $D$ given a particular set of parameters, $P(D)$ is the evidence serving as a normalisation factor, and $P(\theta|D)$ is the posterior distribution. Stochastic techniques such as Markov Chain Monte Carlo (MCMC) are commonly employed to explore the parameter space and reconstruct the posterior efficiently, which encapsulates the intrinsic nature of the alternate scenario under consideration. Using MCMC technique, the MWI model is confronted with observational data within a Bayesian framework. The Table.(\ref{prior}) presents the intervals assigned to the various model parameters, across different scenarios, within which uniform priors are assumed. The combined likelihood  \texttt{Planck18 + lowTT + lowEE + highl CamSpec TTTEEE  +  BICEP/Keck 2018 + sn.pantheon + SH0ES} leverages complementary early and late universe probes, enabling a stringent test of inflationary dynamics, dark energy evolution and the consistency of the inferred Hubble constant.
\begin{table}[htp]
\begin{center}
\begin{tabular}{|c|c|}\hline
Parameters & Priors \\\hline
$H_0$ & [62, 78] \\
$\Omega_b h^2$ & [0.01, 0.023] \\
$\Omega_c h^2$ & [0.1, 0.3]\\
$\tau$ &[0.01, 0.15]\\ \hline
\end{tabular}
\end{center}
\caption{Uniform prior ranges adopted for the parameters appearing in the different models considered in this analysis.}
\label{prior}
\end{table}
In this work, \texttt{Cobaya} is employed in combination with \texttt{CAMB}. The resulting MCMC chains are then analysed using \texttt{GetDist} package in \texttt{Python}. The obtained results are presented in Table.(\ref{mcmctable}). Analysis of the results shows an increase in the value of $H_0$ with increasing $Q$ in MWI, which is higher than that estimated from $\Lambda$CDM  using the same potential $V(\Phi)$ but in the conventional CI . MWI predicts $H_0 \approx $ 71  km s$^{-1}$ Mpc$^{-1}$, whereas, $\Lambda$CDM and CI gives $H_0 \approx 67$  km s$^{-1}$ Mpc$^{-1}$.  Fig.(\ref{mcmc}) compares the contours of the obtained results for MWI with that of CI. Comparison reveals the effect of dissipation from axionic MWI on the cosmological parameters and its role in alleviating the Hubble tension. The $H_0$ estimated from axionic MWI has a broader range of values and is on the higher side compared to CI. Therefore, it is evident that, while CI accounts for the lower value of $H_0$ from the the $\Lambda$CDM based $Planck$ 2018 observation (67.36 $\pm$ 0.54 km s$^{-1}$ Mpc$^{-1}$), the axionic minimal warm inflation can account for the higher values of $H_0$  (74.03 $\pm$ 1.04 km s$^{-1}$ Mpc$^{-1}$) inferred from the local measurements such as $SH0ES$ dataset, thereby pointing towards a possible solution to the Hubble tension. The analysis also shows a small increase in the baryonic density and slightly lower dark matter density (see Table. \ref{mcmctable}) compared to $Planck$ $( \Omega_b h^2=0.0224 \pm 0.0001$ and $\Omega_c h^2=0.120 \pm 0.001)$ \cite{planck}. The decay products from the interactions during warm inflation may be responsible for the increase in baryonic density.\\
\\ 
\begin{table}[h]
\centering
\begin{tabular}{|c|c|c|c|c|c|}
\hline 
Parameters & Q=3 & Q=4 & Q=5 & Q=6 & Q=7 \\
\hline
$H_0$ & 70.98 $\pm$ 0.4173& 71.16 $\pm$ 0.3316 & 71.32 $\pm$ 0.3599 & 71.43 $\pm$ 0.3502 & 71.27 $\pm$ 0.8205 \\
$\Omega_b h^2$ & 0.0228 $\pm$ 0.0001  & 0.0228 $\pm$ 0.00009 & 0.0228 $\pm$ 0.00001 & 0.0228 $\pm$ 0.0001 & 0.0222 $\pm$ 0.0006 \\
$\Omega_ch^2$ & 0.1119 $\pm$ 0.0009 & 0.1115 $\pm$ 0.0007 & 0.1111 $\pm$ 0.0008 & 0.1109 $\pm$ 0.0008 & 0.1169 $\pm$ 0.0032 \\
$\tau$ &0.0588 $\pm$ 0.0027 & 0.0625 $\pm$ 0.0025 & 0.0659 $\pm$ 0.0026 & 0.0685 $\pm$ 0.0026 & 0.0439 $\pm$ 0.0065 \\
\hline 
\end{tabular}
\caption{Marginalised mean and standard deviation of different model parameters obtained from the Bayesian analysis of MWI for various values of the dissipation strength obtained with baseline likelihoods from \texttt{Planck18 + lowTT + lowEE + highl CamSpec TTTEEE  +  BICEP/Keck 2018 + sn.pantheon + SH0ES} datasets.} \label{mcmctable}
\end{table}

\begin{figure}[H]
\begin{center}
\includegraphics[width=0.6\linewidth]{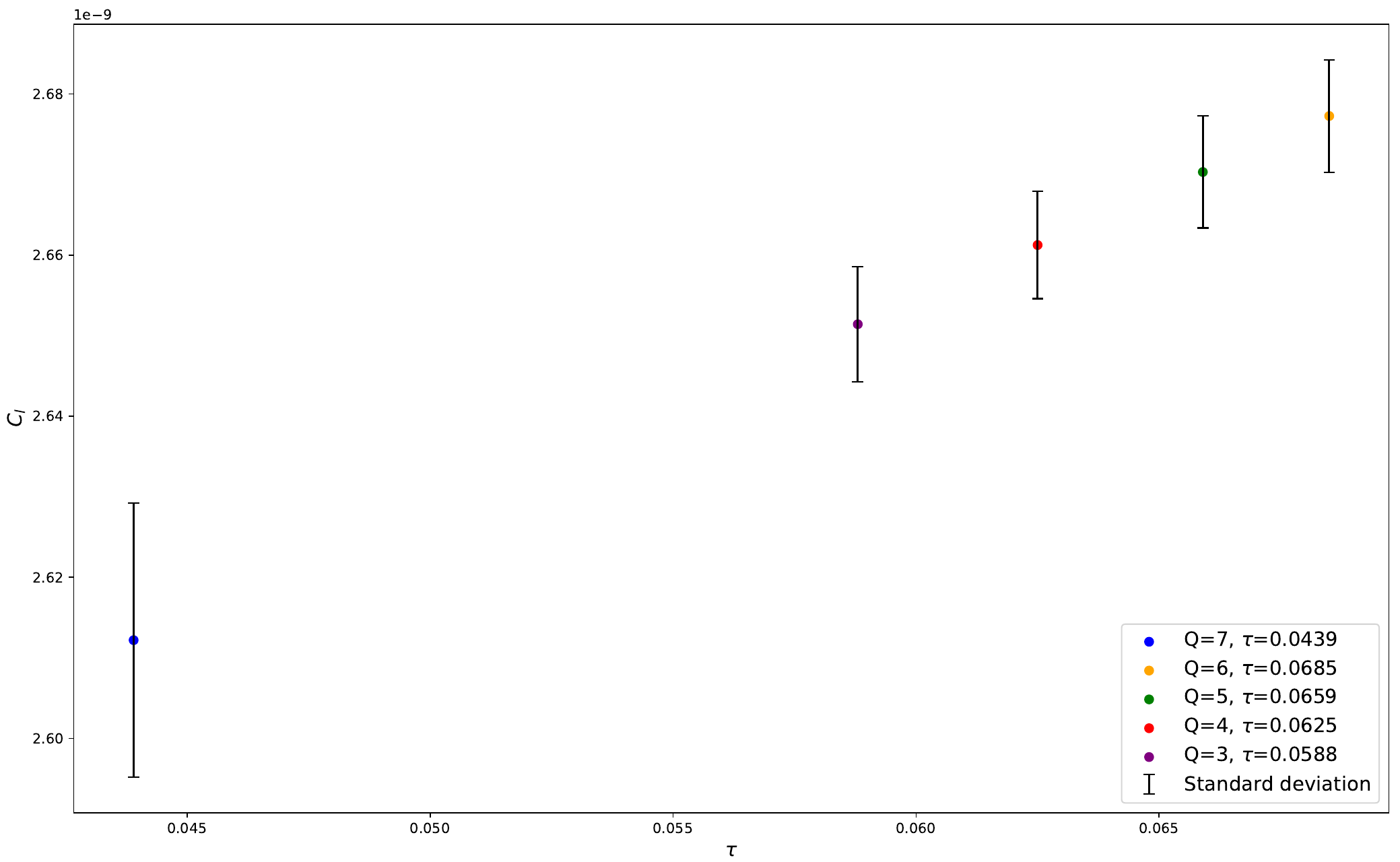}
\caption{Variation of the power of TT mode angular power spectrum ($C_l$) of CMB with the optical depth}
 \label{As}
 \end{center}
 \end{figure}

\noindent The optical depth to reionisation ($\tau$) quantifies the probability that CMB photons were scattered by free electrons once the first stars and galaxies reionised the universe \cite{tauref}. In dynamical dark energy models, the late time expansion history is affected by the evolution of $\omega$. This affects
the mapping between primordial fluctuations and present CMB observables, resulting in difficulties in inferring the growth of structure. Since CMB primarily constrains the combination $C_l=A_s e^{-\tau} (A_s = 2.9 \times 10^{-9}$  is the primordial amplitude)  \cite{planck} , any shift in $\tau$ directly affects the inferred $C_l$. This relation is verified for MWI by taking the bounds of $\tau$ from the MCMC computation for granted (see Table. (\ref{mcmctable})) and the results are presented in Fig. (\ref{As}). The change in the $C_l$ due to increase in $\tau$ is manifested in the TT mode angular power spectrum and is evident from Fig. (\ref{default}). This makes $\tau$ a key parameter in the detection of dynamical dark energy. The results of the present study signal a higher value for $\tau$ (see Table. \ref{mcmctable}) when compared to that from $Planck$ 2018 data ($\tau = 0.054 \pm 0.007$), signalling a dynamical dark energy like behaviour from axionic MWI. In dynamical dark energy models, adjusting $\tau$ can slightly shift the expansion history, thus modifying structure formation. It can influence inferred $H_0$ values and hence plays a non-trivial role in easing Hubble tension. This is clearly noticeable from the obtained results presented in the Table. (\ref{mcmctable}) and Fig. (\ref{mcmc}). This is the first clue towards unifying dynamical EDE and Hubble tension.
  
\subsection{ Matter power spectrum}\label{mps}
\begin{figure}[H]
\begin{center}
\includegraphics[width=0.6\linewidth]{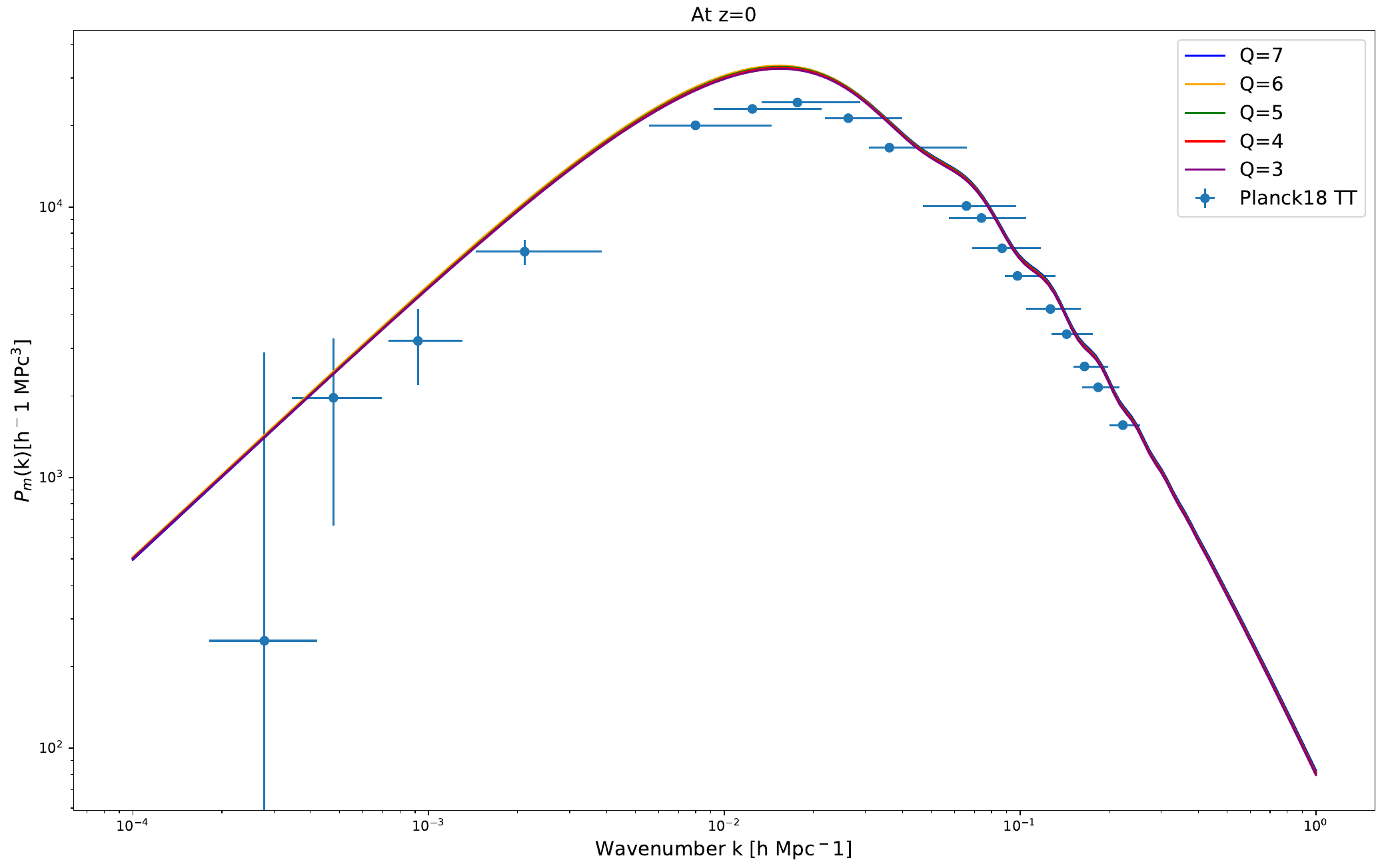}
\caption{Comparison of matter power spectrum for various value of $Q$ from MWI with $Planck$18 data.}
\label{pk}
\end{center}
\end{figure}

The matter power spectrum encodes the growth of density perturbations. The modified friction term in the perturbation equation of warm inflation (see Eq. (\ref{motion})) can affect the growth of structures, and, if detected, could be considered a signature of dynamical dark energy. To differentiate genuine dark energy evolution from simple parameter shifts within the $\Lambda$CDM framework, the matter power spectrum $P_M(k)$ corresponding to various dissipative strengths of MWI (see Fig.(\ref{pk})) is generated using \texttt{CAMB}. Depending on whether dark energy becomes stronger or weaker in the past, there will be suppression or enhancement of late time clustering.
 Thus, dynamical dark energy can induce changes in the matter-radiation equality and in the expansion history. Typically, a leading dark energy at earlier times can suppress the structure growth, which can lower the power at small scales. However, the results of the current study show that dynamical dark energy from MWI is weaker, leading to enhanced growth and higher clustering amplitude compared to that from $Planck 18$ (see Fig. \ref{pk}). Therefore, the results confirm the dynamical nature of early dark energy from axionic MWI.
\subsection{Dynamial dark energy and Hubble tension}\label{ddeht}
\begin{figure}[H]
\centering
\includegraphics[width=0.6\linewidth]{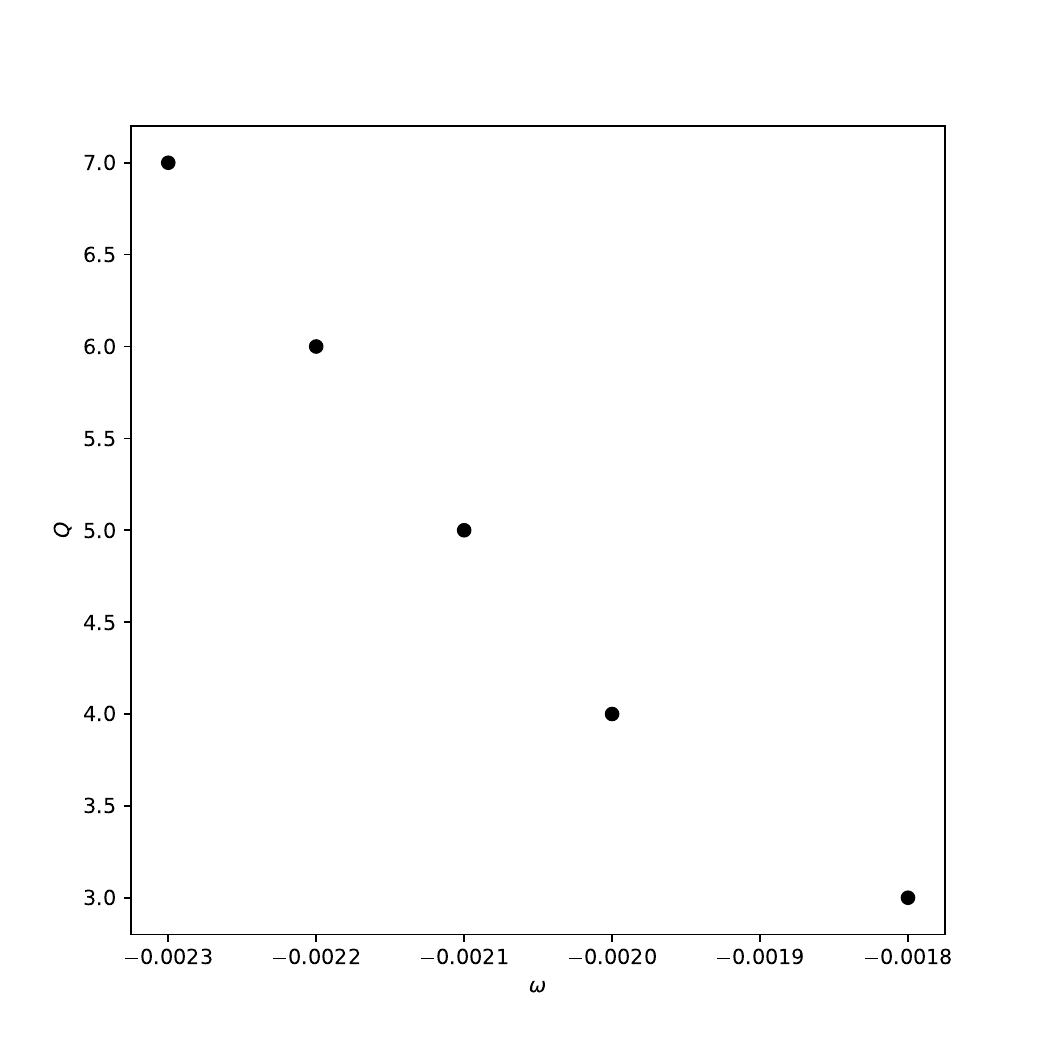}
\caption{\label{Qw}Evolution of the dissipative strength ($Q$) of MWI with the equation of state ($\omega$) of early dark energy.}
\end{figure}

\begin{figure}[H]
\begin{center}
\makebox[\linewidth]
{
   \begin{subfigure}{0.8\linewidth}
    \includegraphics[width=\linewidth]{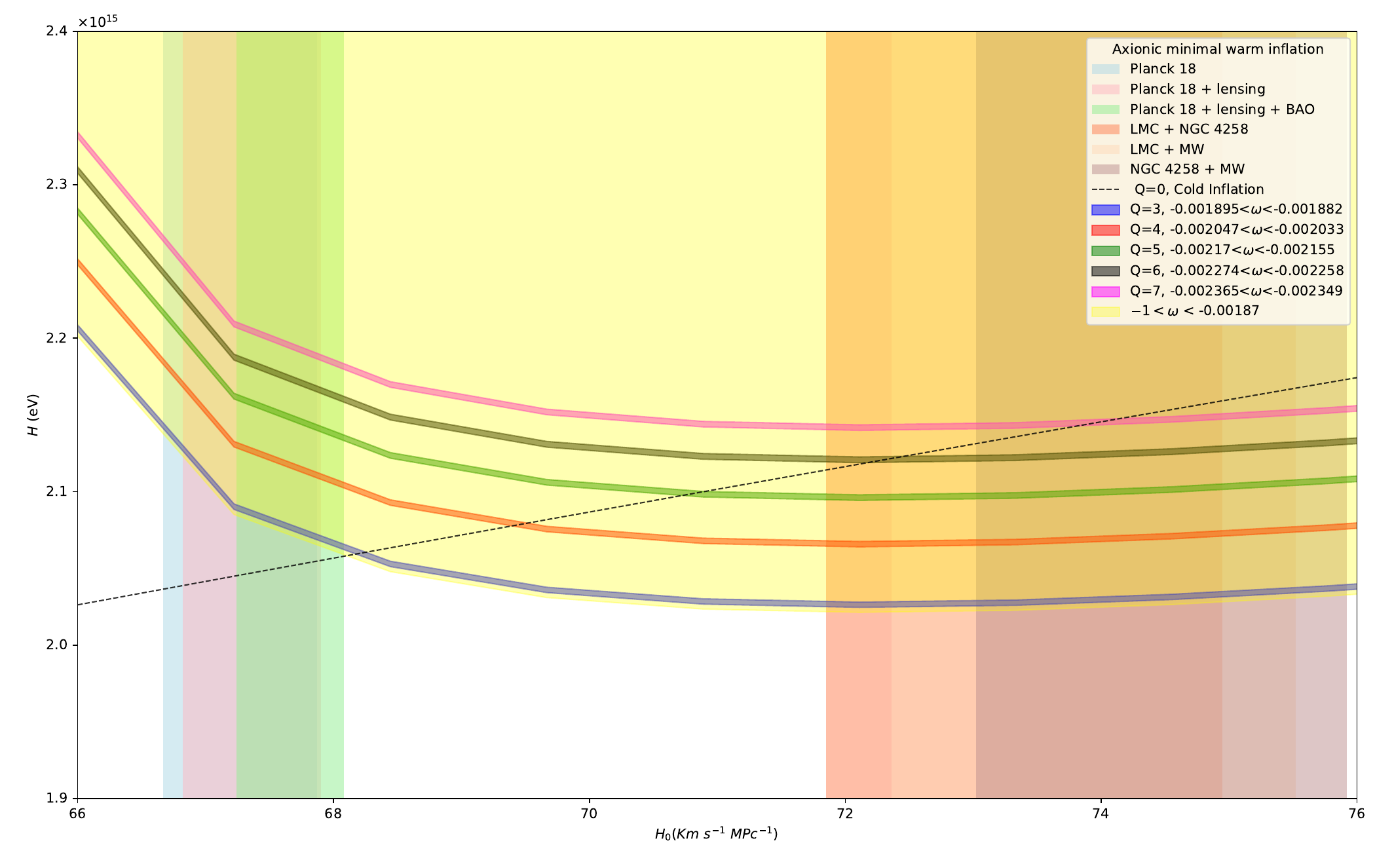}
     \caption{}
     \end{subfigure}
     }\par 
     \makebox[\linewidth]{
     \begin{subfigure}{0.8\linewidth}
    \includegraphics[width=\linewidth]{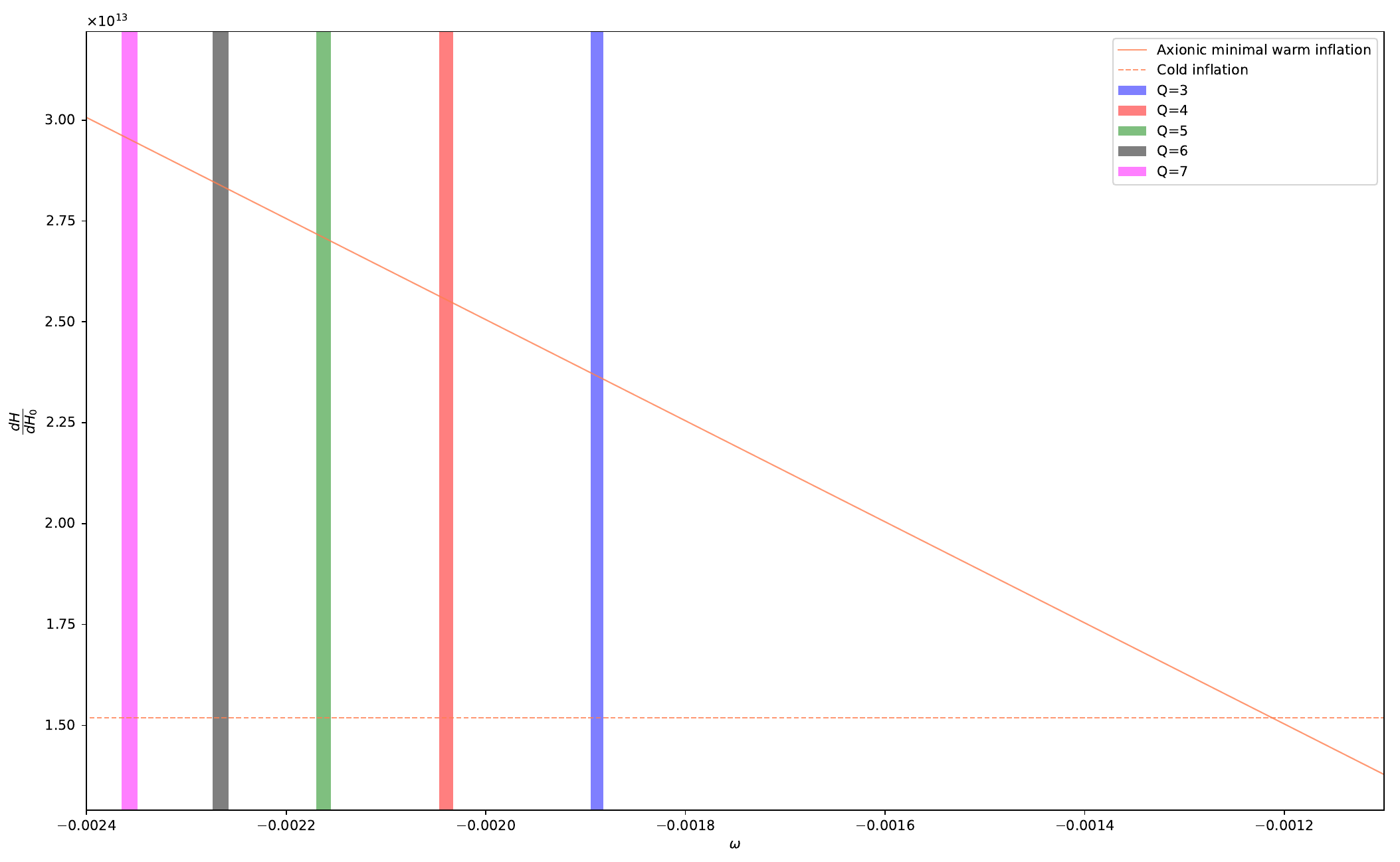}
    \caption{}
     \end{subfigure}
     } \par
     \end{center}
\caption{\label{overallHT} (a) Dynamical dark energy effect from MWI on the Hubble parameter during inflation $(H)$ for a range of present Hubble parameter $(H_0)$ with $Planck18$ and $SH0ES$. (b) The effect of $\omega$ on the rate of change of Hubble parameter $( \frac{dH}{dH_0})$ in MWI compared with CI. }
\end{figure}

\begin{figure}[H]
\centering
 \includegraphics[width=0.6\linewidth]{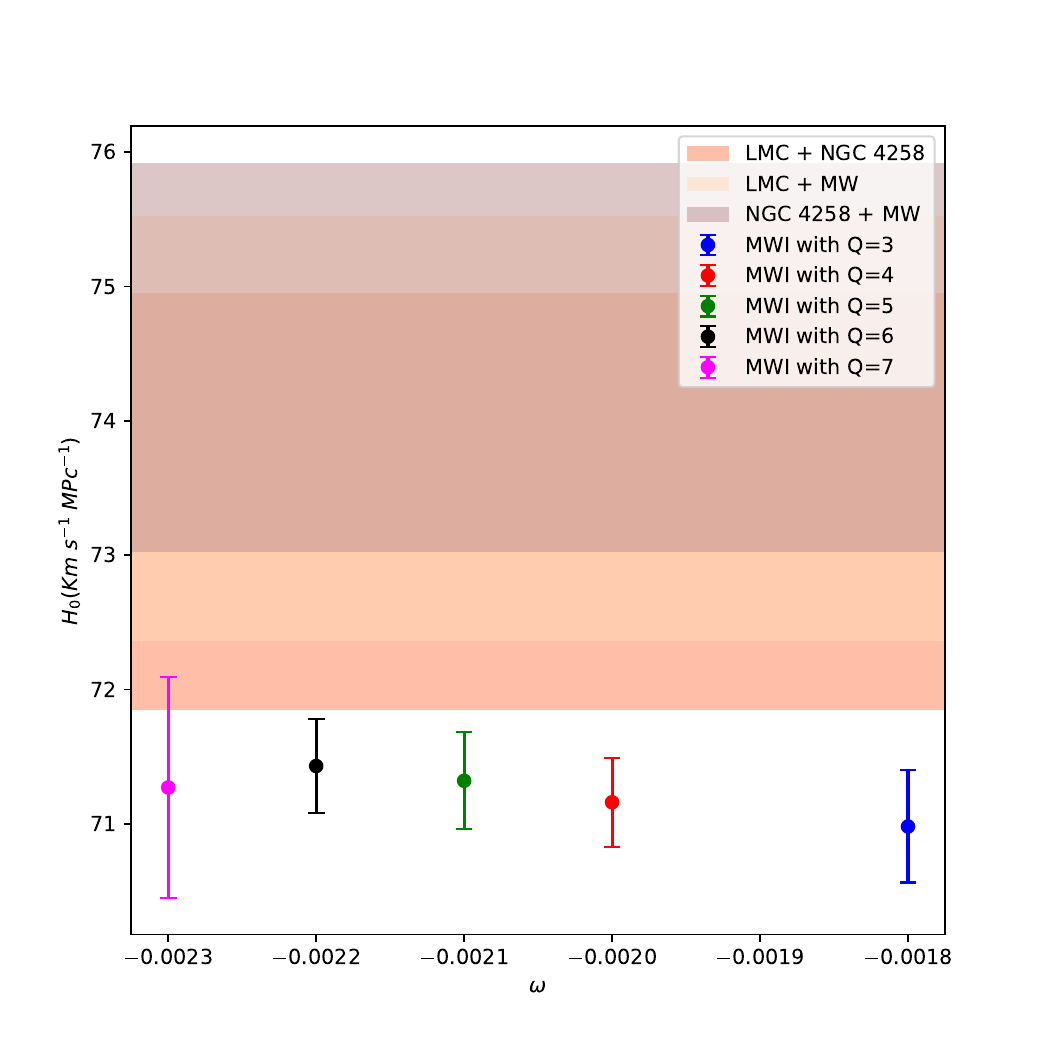}
\caption{\label{H0} The variation of $H_0$ for a range of $\omega$ obtained from various $Q$ in MWI compared with $SH0ES$.  }
\end{figure}

Although the $\Lambda$CDM model is known as the concordance model in cosmology, the fundamental nature of the cosmological constant ($\Lambda$) and dark matter remains unknown. Despite its success in explaining most of the the observations in cosmology, it faces several challenges \cite{c1, c2, c3, c4}. Recent improvements in observational precision have exposed several tensions among different datasets, the most prominent of which is the Hubble tension, which continues to pose a major challenge for cosmologists \cite{kami, planck, shoes, ldl, devalentino}. Even though a range of possible systematic uncertainties has been explored, no broad consensus has emerged regarding their ability to fully account for the discrepancy \cite{chl, model, odi, dv1, tale, anu, bour}. Given the relative simplicity and robustness of local measurements of the Hubble constant, many proposed resolutions invoke new physics to reconcile the value inferred from CMB observations. From this standpoint, the tension may be viewed as a potential signal of physics beyond the standard $\Lambda$CDM paradigm. Since the $\Lambda$CDM model is based on a flat FLRW universe, studies focusing on the mathematical side of general relativity influenced by the interaction between matter and non-linear geometry of spacetime \cite{35, nf, ct, ts} are currently being explored to address these tensions. But on the other side, early universe solutions are favoured at present \cite{pedro, es, es1, es2, es3, es4, es5, es5, es7, es8, es9}, with EDE being a feasible example.
 Given the robustness of local measurements of the Hubble constant, most proposed resolutions focus on introducing new physics that raises the value of $H_0$ inferred from CMB observations. A direct approach for raising the Hubble constant inferred from CMB data is to introduce a small modification to the early universe dynamics, while leaving the late time evolution essentially unchanged.\\
\\
Therefore, in this paper, the effect of dynamical early dark energy naturally generated during MWI on the present Hubble parameter $H_0$ is explored. In a warm inflationary setting, the inflaton potential can be expressed in terms of the dissipation coefficient and the temperature of the thermal bath during warm inflation from Eqs (\ref{V}), (\ref{fried}) and (\ref{newp}) as,
\begin{equation}\label{st}
\frac{m^2 \sqrt{\gamma H}T}{4 \pi^2} = \frac{\dot \Phi^2}{2} -  3\omega m_{pl}^2\Omega_\Lambda H_0^2.\\
\end{equation}
Eq. (\ref{st}) can be modified further to incorporate the warm inflationary artefacts by
substituting for $\dot \Phi$ from Eq. (\ref{eqwarm}), 
\begin{equation}\label{stnew}
\frac{m^2 \sqrt{\gamma H}T}{4 \pi^2} \bigg( 1 - \frac{m^2}{9H^2(1+Q)^2} \bigg) = -3\omega m_{pl}^2\Omega_\Lambda H_0^2.
\end{equation}
Substituting for the ratio $\frac{T}{H}$ from Eq. (\ref{TH}) and taking $m \sim H$, $\gamma=3QH$, Eq. (\ref{stnew}) takes the form,
\begin{equation}\label{nchaos}
\frac{c\sqrt{3Q}H^4}{4\pi^2} \bigg[\frac{Q}{(1+Q)^2}\bigg]^{\frac{1}{4}} \bigg(\frac{1}{9Q^2}-1\bigg)= 3\omega m_{pl}^2 \Omega_{\Lambda} H_0^2.
\end{equation}
Strong dissipation implies $Q>>1$, therefore $1+Q \approx Q$, rearranging Eq. (\ref{nchaos}), a relation between the Hubble parameter during inflation ($H$) and the present Hubble parameter ($H_0$) can be obtained in terms of $Q$ and $\omega$ as,
\begin{equation}\label{bj21}
H = \bigg[ 12 \pi\Omega_\Lambda \frac{\omega}{\frac{\sqrt{3}c}{\pi}Q^{\frac{1}{4}}\bigg(\frac{1}{9Q^2}-1\bigg)} m_{pl}^2 \bigg]^{\frac{1}{4}} H_0^{\frac{1}{2}}.
\end{equation}
\\
Similarly, the relation between $H$ and $H_0$ for a quadratic inflationary potential in axionic MWI is obtained in Ref. \cite{cqg2} under the same conditions and interactions,
\begin{eqnarray}\label{quad}
H= \Bigg[ \frac{12 \pi \Omega_\Lambda}{1+\frac{\sqrt{3}cQ^\frac{1}{4}}{\pi}} \Bigg]^\frac{1}{4} \sqrt{m_{pl}H_0}.
\end{eqnarray}
Therefore, equating Eq. (\ref{bj21}) and Eq. (\ref{quad}), 
the value of $\omega$ corresponding to each possible $Q$ in MWI can be obtained,
\begin{equation}\label{w}
\omega = \frac{\frac{\sqrt{3}c}{\pi}Q^{\frac{1}{4}}\bigg(\frac{1}{9Q^2}-1\bigg)}{1+ \frac{\sqrt{3}c}{\pi}Q^{\frac{1}{4}}}.
\end{equation}
Eq. (\ref{w}) is studied for various possible range of values of $Q$ and the results are presented in Table. (\ref{HWQ}). $Q$ arises from the interactions between the axionic inflaton field and the non Abelian gauge field during warm inflation.
Since the axionic MWI lies in a strong dissipation regime, substituting $Q>1$ in Eq. (\ref{w}) gives a negative value for the equation of state. This negative pressure suggests that early dark energy is inherent in warm inflation. The evolution of the negative equation of state with $Q$ is the second clue towards dynamical dark energy. Analysis shows that the beginning of warm inflation (when the dissipation is more than expansion $Q=7$) yields lower value of equation state ($\omega = -0.0023$). As the warm inflation proceeds, $Q$ decreases, $\omega$ increases and when the thermal bath falls out of thermal equilibrium ($\gamma \neq 3QH$) completely,  $\omega$ reaches $-0.0018$ marking the end of inflation. 
Since $Q$ and $\omega$ are mutually dependent, conversely, it can also be argued that $\omega$ drives the dynamics of minimal warm inflation (see  Fig. (\ref{Qw})). Or in other words, dynamical dark energy is embodied in warm inflation.
\begin{table}[h]
\begin{center}
\begin{tabular}{|c|c|}
\hline
Q & $\omega$ \\ \hline 
7 &-0.0023 \\
6 &-0.0022 \\
5 &-0.0021  \\
4 &-0.0020  \\
3 &-0.0018  \\
\hline 
\end{tabular}
\caption{\label{HWQ}Estimate of the equation of state of dark energy from MWI for various $Q$ using Eq. (\ref{w})}
\end{center}
\end{table}%
\noindent Using Eq. (\ref{bj21}), the effect of varying dissipation strength and dynamical dark energy from MWI on $H_0$ is studied. The results are presented in Fig. (\ref{HT} (a), (c), (e), (g) and (i)).  A close scrutiny reveals that as the warm inflation progresses, the Hubble parameter during inflation ($H$) decreases rapidly with increase in $\omega$. In order to distinguish the effect of dynamical dark energy on the Hubble parameter, the fractional change in the Hubble parameter ($\frac{dH}{dH_0}$) can be computed
\begin{equation}\label{seq}
\frac{dH}{dH_0} = \bigg[ 12 \pi\Omega_\Lambda \frac{\omega}{\frac{\sqrt{3}c}{\pi}Q^{\frac{1}{4}}\bigg(\frac{1}{9Q^2}-1\bigg)} m_{pl}^2 \bigg]^{\frac{1}{4}}  \frac{1}{2} H_0^{-\frac{1}{2}}.
\end{equation}
Eq. (\ref{seq}) can be used to examine the rate of change of the Hubble parameter with equation of state ($\omega$), and the obtained results are presented in Fig. (\ref{HT} (b), (d), (f), (h) and (j)). It is interesting to observe that $\frac{dH}{dH_0}$ decreases drastically with $\omega$. This can be interpreted as either a decrease in $dH$ or as an increase in $dH_0$. While the former argument is more elusive, the latter is in accordance with the results obtained in the previous section. Fig. (\ref{overallHT}(a)) consolidates the results obtained from Eq. (\ref{bj21}) and compares it with various $Planck$18 and $SH0ES$ bounds on $H_0$. There is a stark difference between the Hubble parameter 
 predicted in cold inflation (CI) and warm inflation (WI), and this difference becomes more pronounced as the dissipative contribution and hence the contribution from dynamical dark energy increases. In the cold inflationary limit ($Q=0$), the inferred value of $H_0$ aligns with that obtained within $\Lambda$CDM, whereas allowing for dissipation ($Q\neq0$)) in warm inflation can shift $H_0$ toward the higher values indicated by supernova observations. In this way, both the lower and higher estimates of $H_0$ can, in principle, be accommodated within a unified framework of dynamical dark energy and WI. As the warm inflation approaches its last stages, $T>H$ is no longer satisfied in the thermal bath, and the fractional change in the Hubble parameter during warm inflation approaches CI and coincides at higher $\omega$ (see Fig. (\ref{overallHT}(b))). Further, the values of $H_0$ obtained from the MCMC analysis (see Table. (\ref{mcmctable})) are studied with their corresponding equation of state obtained from Eq. (\ref{w}). The results are compared with $SH0ES$ data and are presented in Fig. (\ref{H0}). The analysis shows that $H_0$ increases with $\omega$ and approaches the late time values of $H_0$ estimated by $SH0ES$ team at higher values of $\omega$. Also, the range of $\omega$ required for obtaining the increased $H_0$ also increases with $Q$. Therefore, dissipation from MWI plays a role in the evolution of $\omega$ which in turn may be responsible for the increased $H_0$ at late times. The expressions in Eqs. (\ref{bj21}, \ref{w} and \ref{seq}) are the main results of this paper that substantiate the role of dynamical dark energy from axionic minimal warm inflation in resolving Hubble tension (see Figs. \ref{Qw}, \ref{HT}, \ref{overallHT} and \ref{H0}).\\
\\
Thus, a viable warm inflationary scenario may effectively behave like an evolving early dark energy component, offering an alternate path to reconcile the early universe  $H_0$ inferred from CMB measurements by the $Planck$ Collaboration with the late time value reported by the $SH0ES$ team.  From this perspective, the Hubble tension could reflect the imprint of thermal fluctuations generated during warm inflation, which may contribute to the dynamical nature of dark energy. Recent studies \cite{eede, eede1, eede2, eede3, eede4, eede5, eede6, eede7, eede8, eede9, eede10, eede11, eede12} have highlighted the importance of dark energy dynamics and its broader role in addressing major cosmological tensions, thus supporting the results of the present study. 
Furthermore, recent results from DESI DR2, which support effective field theory formulations of dark energy and quantum gravity inspired extensions \cite{1, 2, 3, 4}, provide additional insights to this work. Given that quantum gravity effect has also been suggested as a possible mechanism for alleviating the Hubble tension \cite{cqg2, anu}, the findings of the present study further strengthen the existence of a quantum gravity phase preceding warm inflation. Thus, an additional merit of this work lies in its potential to validate quantum gravity. The generation of thermal gravitational waves from WI is studied in Ref. \cite{tgw}. Testing primordial gravitational waves and quantum gravity within the framework of warm inflation would further enhance the significance of this study. However, a thorough investigation along these lines lies beyond the scope of the present work.
\begin{figure}[H]
\centering
\makebox[\linewidth]
{
   \begin{subfigure}{0.40\linewidth}
    \includegraphics[width=\linewidth]{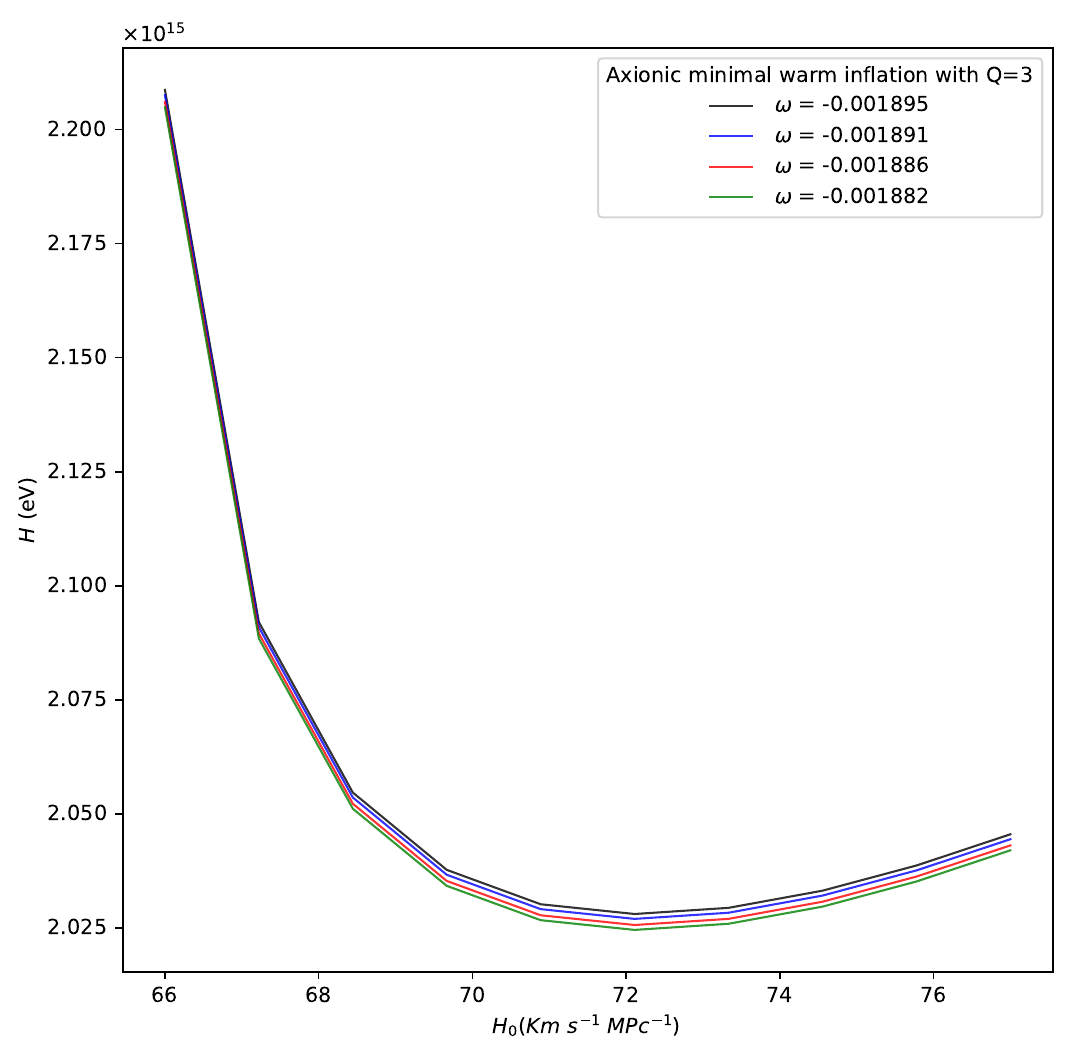}
     \caption{}
     \end{subfigure}
     \begin{subfigure}{0.40\linewidth}
    \includegraphics[width=\linewidth]{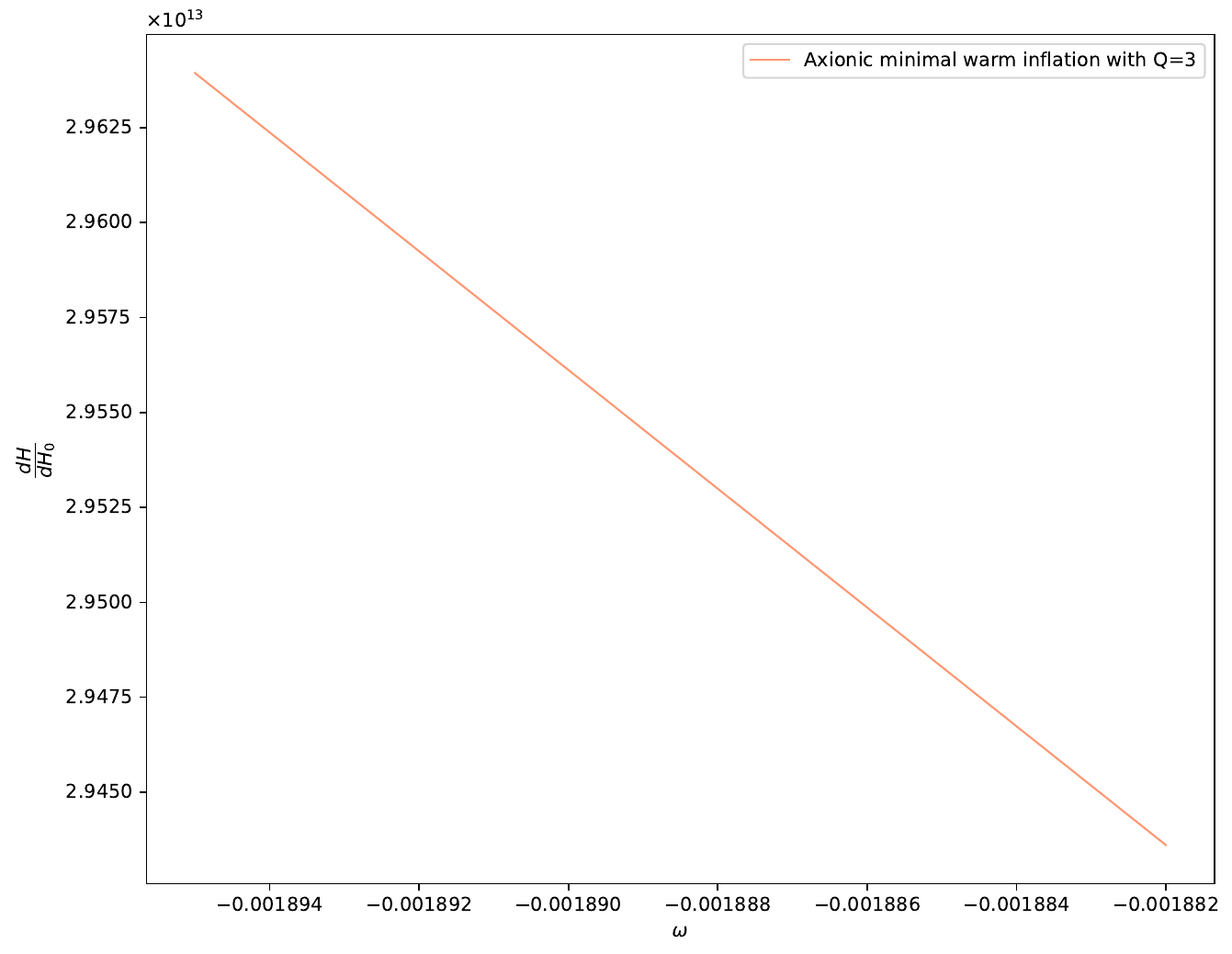}
    \caption{}
     \end{subfigure}
}\par
\makebox[\linewidth]
{
   \begin{subfigure}{0.40\linewidth}
    \includegraphics[width=\linewidth]{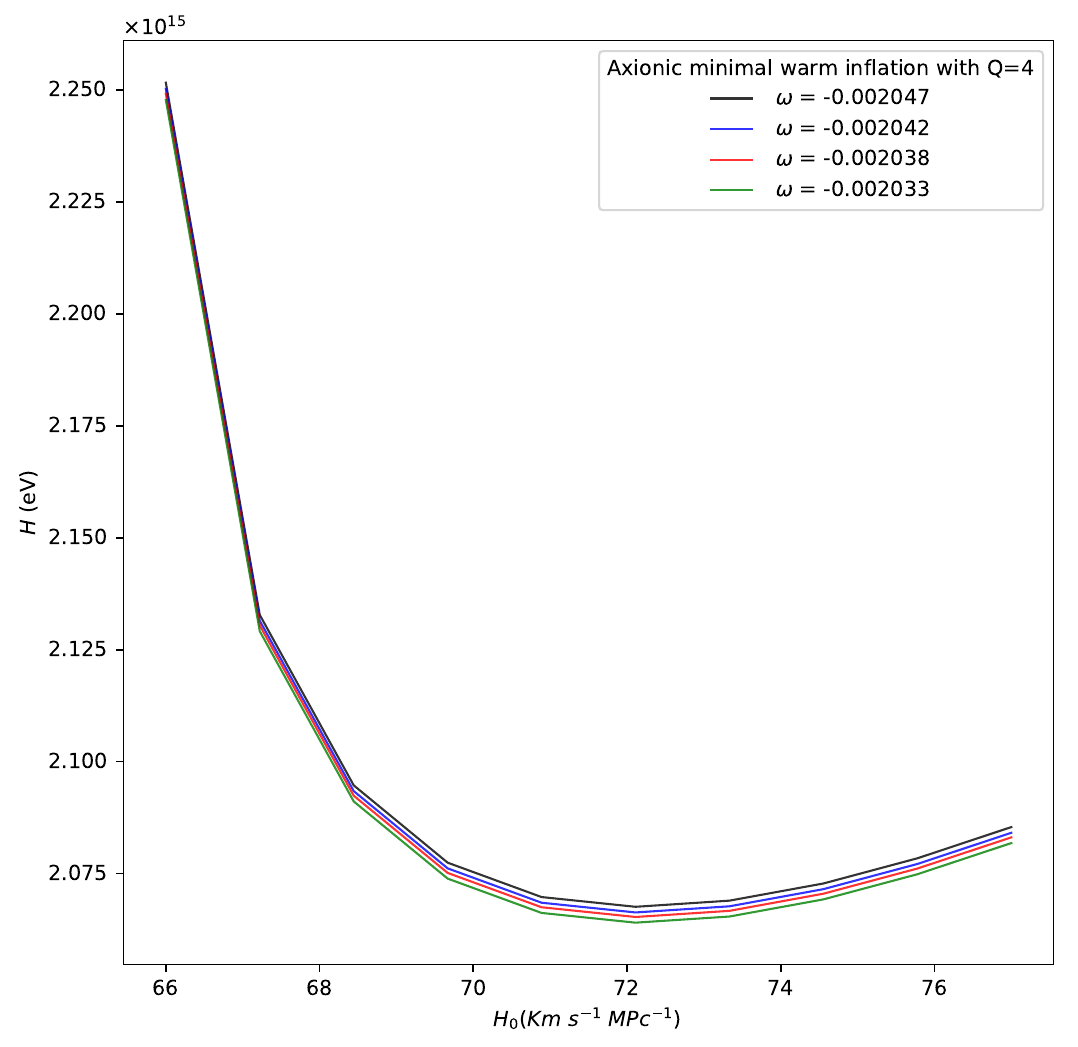}
     \caption{}
     \end{subfigure}
     \begin{subfigure}{0.40\linewidth}
    \includegraphics[width=\linewidth]{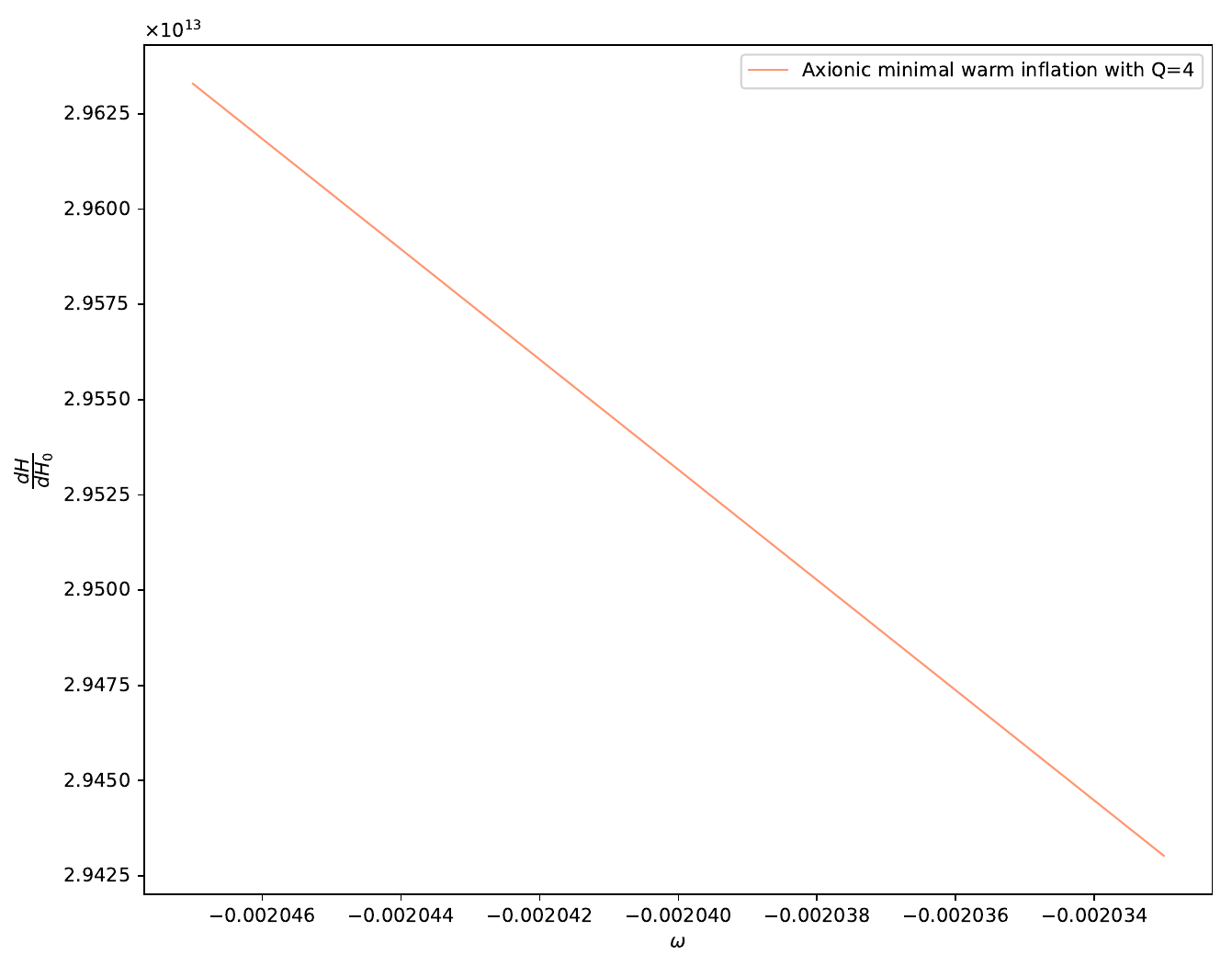}
    \caption{}
     \end{subfigure}
}\par
\makebox[\linewidth]
{
   \begin{subfigure}{0.40\linewidth}
    \includegraphics[width=\linewidth]{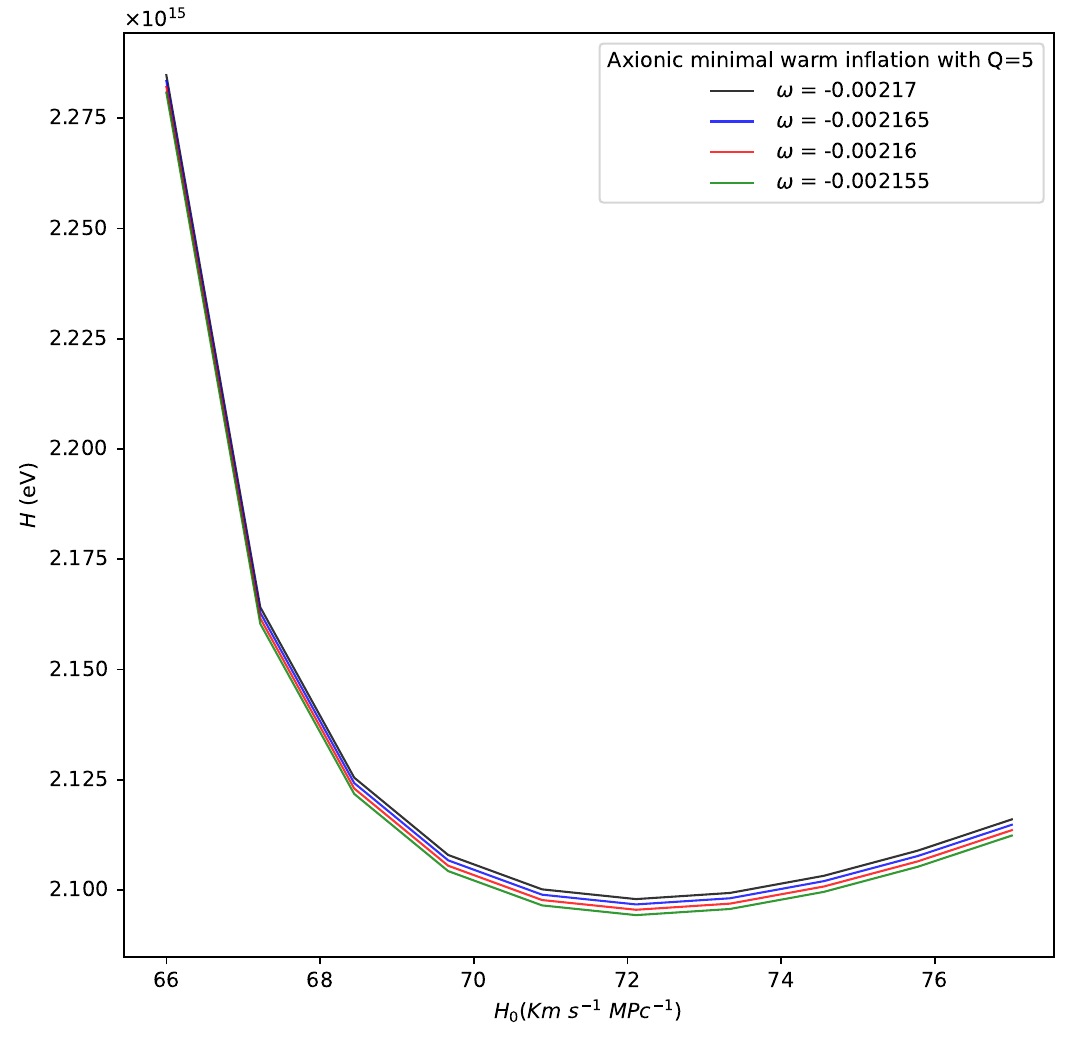}
     \caption{}
     \end{subfigure}
     \begin{subfigure}{0.40\linewidth}
    \includegraphics[width=\linewidth]{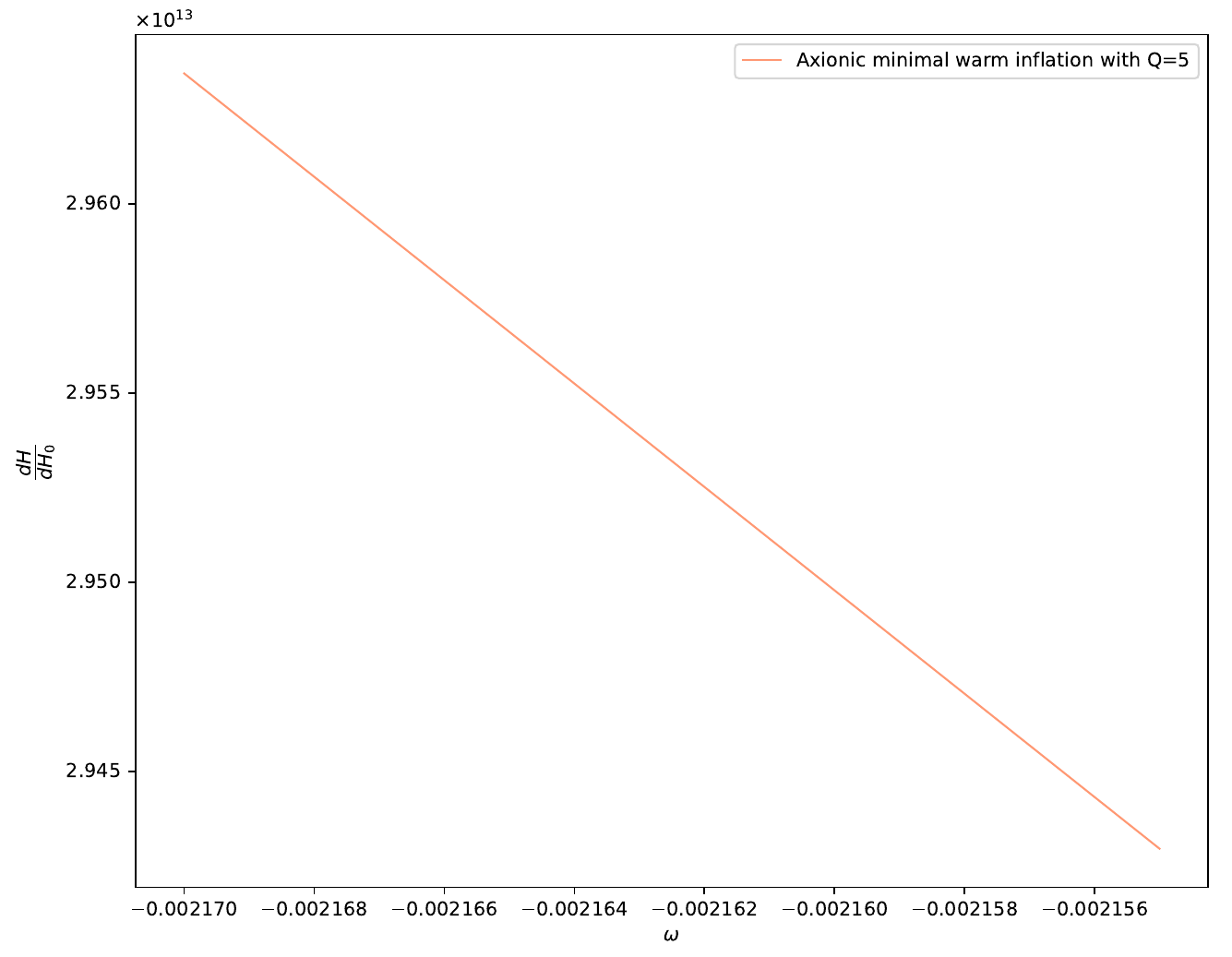}
    \caption{}
     \end{subfigure}
}\par
\end{figure}%

\begin{figure}[H]\ContinuedFloat
\makebox[\linewidth]
{
   \begin{subfigure}{0.40\linewidth}

    \includegraphics[width=\linewidth]{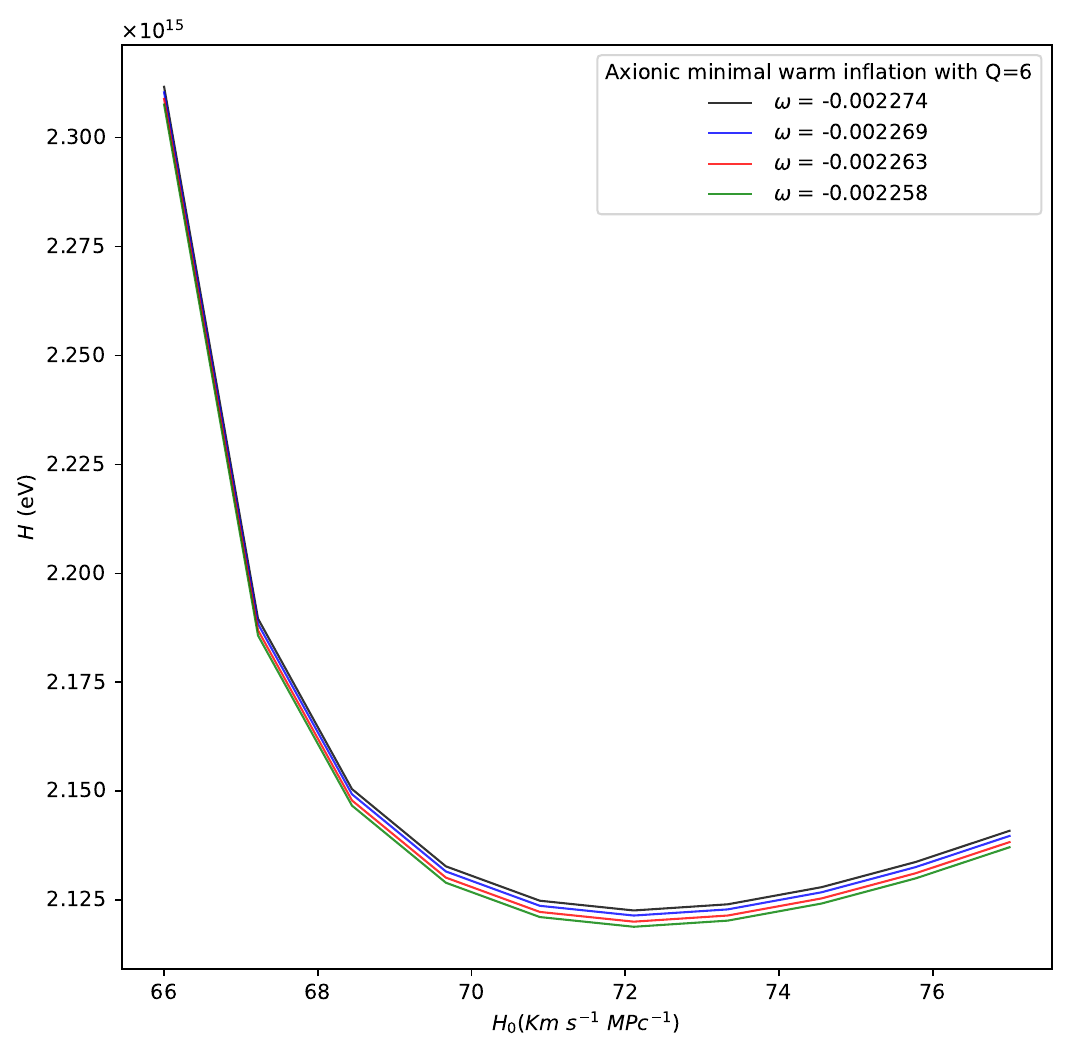}
     \caption{}
     \end{subfigure}
     \begin{subfigure}{0.40\linewidth}
    \includegraphics[width=\linewidth]{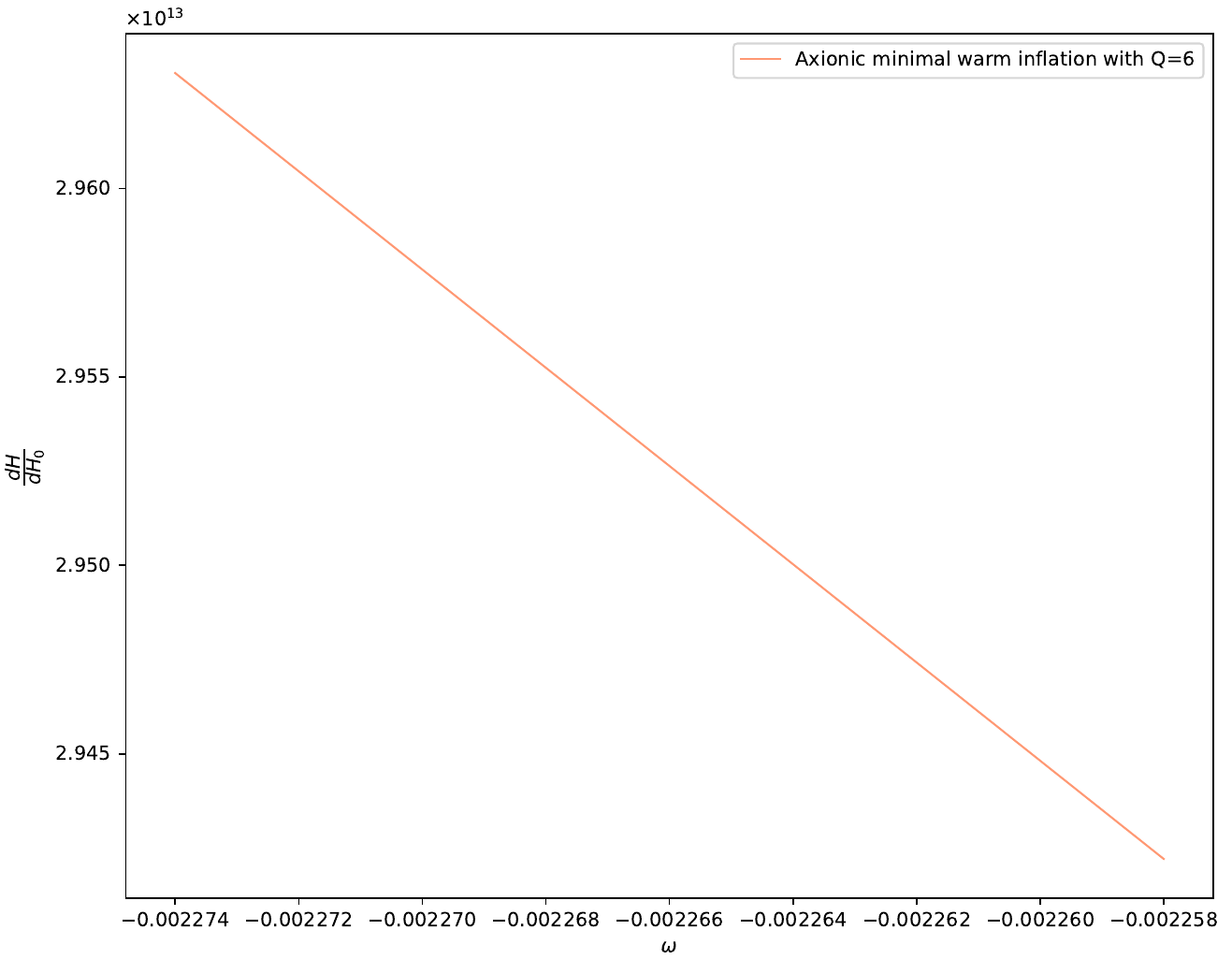}
    \caption{}
     \end{subfigure}
}\par 

\makebox[\linewidth]
{
   \begin{subfigure}{0.40\linewidth}
    \includegraphics[width=\linewidth]{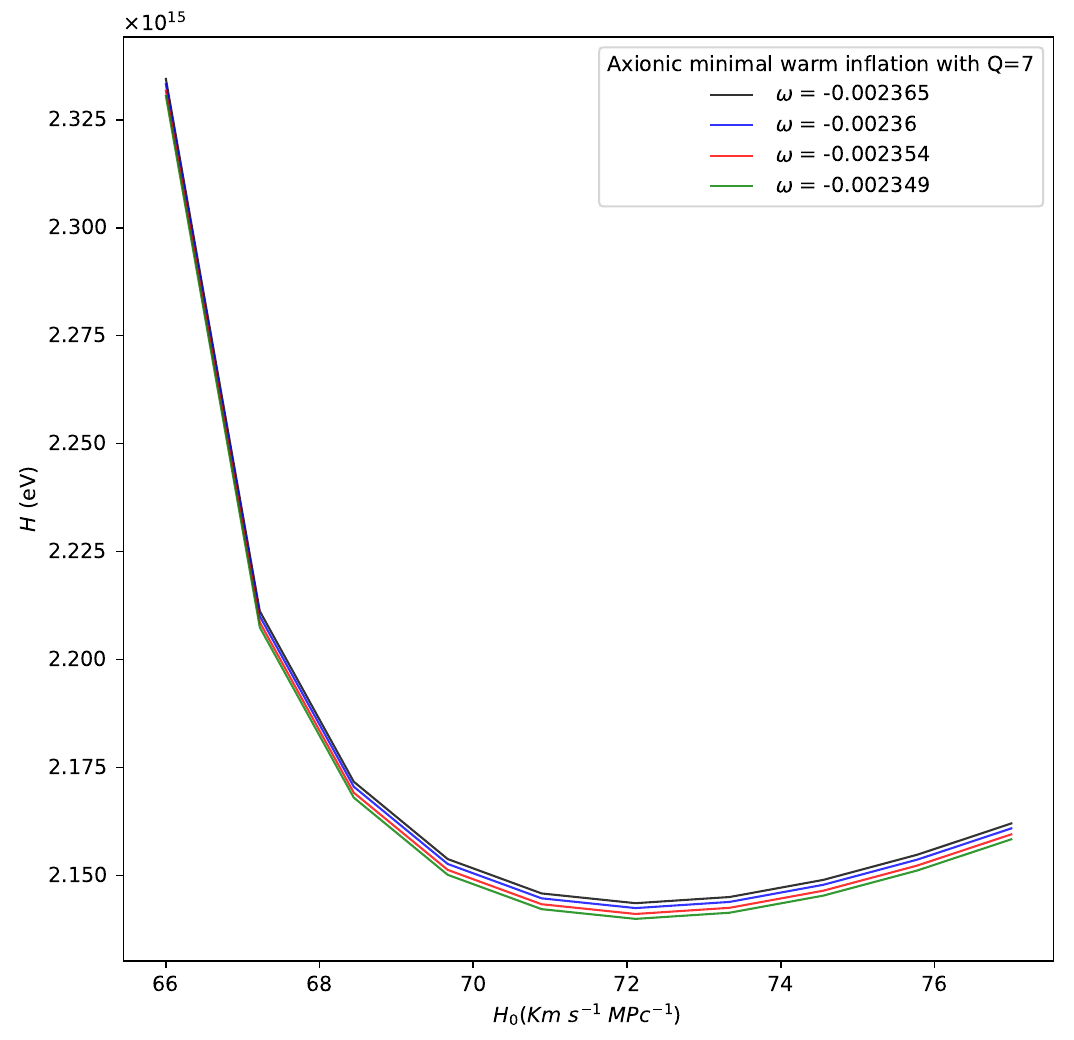}
     \caption{}
     \end{subfigure}
     \begin{subfigure}{0.40\linewidth}
    \includegraphics[width=\linewidth]{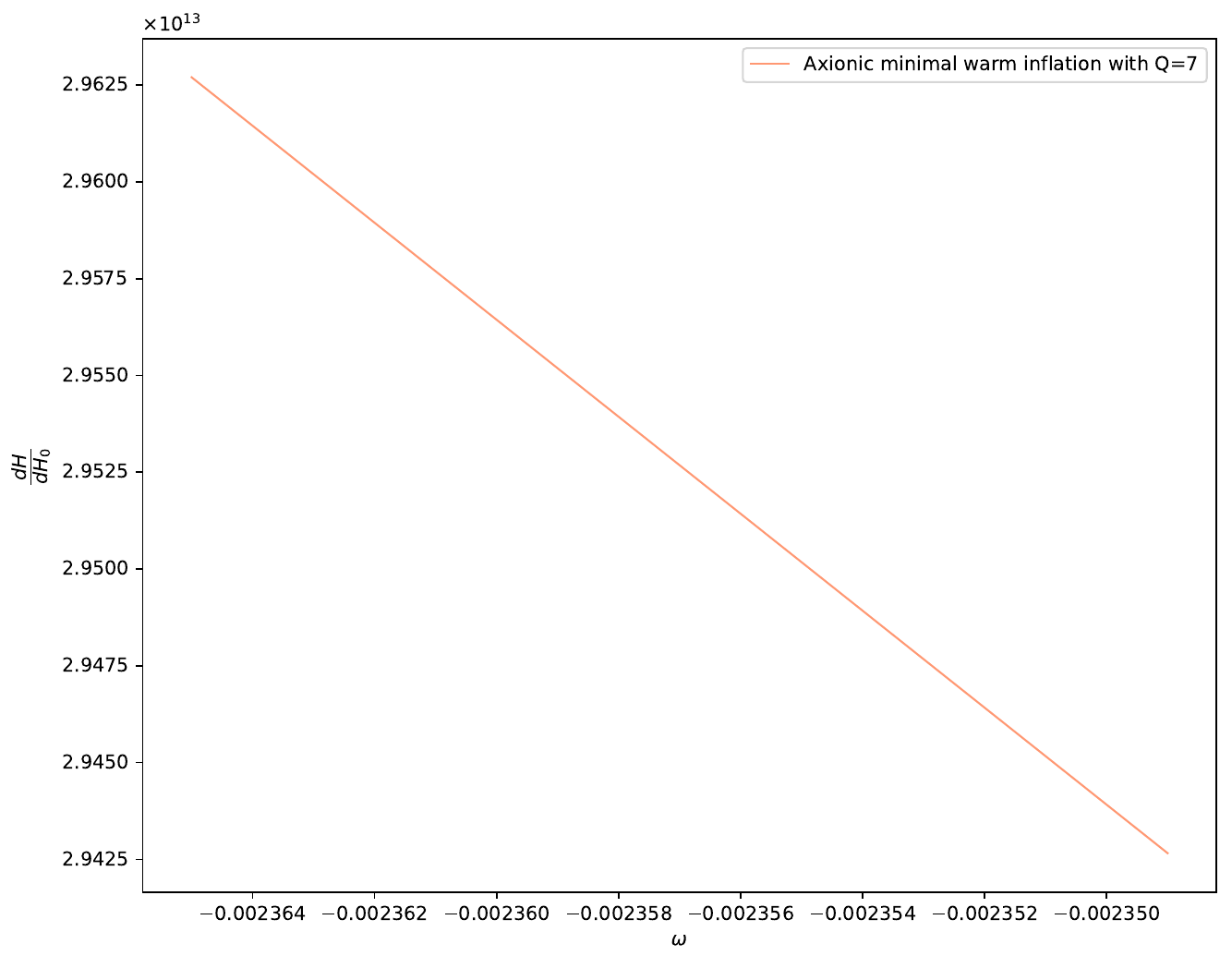}
    \caption{}
     \end{subfigure}
}\par
\caption{\label{HT}  Dynamical dark energy effect on the Hubble parameter during inflation $(H)$ for a range of present Hubble parameter $(H_0)$ for various $Q$ in MWI  (left panel). The effect of $\omega$ on the rate of change of Hubble parameter $( \frac{dH}{dH_0})$ for various $Q$ in MWI (right panel).}
\end{figure}
\section{Conclusions}\label{conc}
In this work, a quadratic axionic potential coupled to non Abelian gauge field is investigated in a warm inflationary framework. The theoretical TT mode angular power spectrum of CMB for MWI hints at the dynamical nature of dark energy. The characteristic features of the angular power spectrum is determined by the specific details of the microphysics of the warm inflationary model under consideration. 
It can be concluded that the thermal bath in MWI behaves like an exotic energy, acting as dark energy with varying equation of state at early times, implying the existence of dynamical dark energy. Therefore, WI supports $\omega \neq -1$, significantly deviating from the standard $\Lambda$CDM model parameter ($\omega=-1$) complementing the result of baryon acoustic oscillation measurements from DESI DR2.  Unlike the other phenomenological models of dark energy, the dark energy component in this approach derives the dynamical effect naturally from the first principles of the warm inflationary set up.
The interactions in the multi field WI that give rise to dynamical dark energy may inturn be responsible for the increase in the expansion rate of the early universe. More precisely, as the axionic field decays, it can behave like dynamical  early dark energy. Axion in MWI is just a representative candidate from the known sector. The involvement of multiple fields in WI may support particles from hidden sectors as well as any new species thus proposing an increased cosmic budget. The possibility of interaction between dark matter and dark energy in WI cannot be completely ignored. In this regard, the work also offers potential insight into the dark universe and encourages further model building in WI by including evolving dark energy. The various models of WI offer a theoretical foundation to explain the evolution of early dark energy as well as the interaction between dark energy and dark matter. The relation between MWI, dynamical dark energy, expansion rate and the cosmological observables such as scalar spectral index and tensor to scalar ratio highlights the relevance of cosmological surveys in understanding the early universe through inflation. Thus, the current work has overcome the observational challenges in detecting dynamical dark energy through CMB. Subsequent outcomes of this study reinforce the existence of dynamical dark energy in the early universe, posing several challenges to $\Lambda$CDM, including a complete revision of the parameter space. The unified framework incorporating dynamical dark energy and warm inflation provides a mechanism to reconcile the early and late time estimates of $H_0$. 
Therefore, MWI promises a fundamental framework for dynamical dark energy that may explain the increased expansion rate of the early universe, thereby alleviating one of the major tensions in cosmology. Beyond general relativity, inflation and particle interaction, this can also lead to a quantum gravity phase in the early universe. This work is of interest not just to cosmology but also to physics and science in general. 
\section*{Acknowledgments}
AB acknowledges the financial support of Prime Minister’s Research Fellowship (PMRF ID: 3702550) provided by the Ministry of Education, Government of India.

\end{document}